\documentclass[aps,prd,11pt,superscriptaddress,notitlepage,longbibliography,nofootinbib,tightenlines,groupedaddress]{revtex4-1}

\usepackage{amsmath,amssymb,amsfonts}
\usepackage{graphicx}
\usepackage{subfigure}
\usepackage{color}
\usepackage{hyperref}
\usepackage{dsfont}
\usepackage{slashed}
\definecolor{darkred}{rgb}{0.8,0.1,0.1}
\hypersetup{colorlinks=true, linkcolor=darkred, citecolor=blue, linktoc=page}

\newcommand{\cc}{\text{c.c.}}
\newcommand{\Dnr}{\ensuremath{\slashed{\mathcal D}}}

\newcommand{\Pg}[2]{P_{#1}^{\hphantom{#1}#2}}
\newcommand{\Pt}{\ensuremath{\tilde {\mathcal P}}}

\makeatletter\def\l@subsubsection#1#2{}%
\makeatletter\def\l@subsection#1#2{}%
\makeatother

\usepackage[Symbol]{upgreek}

\begin{document}
\title{Spinor fields in general Newton-Cartan backgrounds}

 \author{John F.~Fuini III}
 \email{fuini@uw.edu}
 \author{Andreas Karch}
 \email{akarch@uw.edu}
 \author{Christoph F.~Uhlemann}
 \email{uhlemann@uw.edu}
 \affiliation{Department of Physics, University of Washington, Seattle, WA 98195-1560, USA}

\begin{abstract}
We give a covariant construction of Lagrangians for spinor fields in generic Newton-Cartan backgrounds.
A non-relativistic Dirac/L\'evy-Leblond operator and the associated fields are obtained from relativistic analogues by a limiting procedure.
The relativistic symmetries induce the complete set of non-relativistic symmetries, including Milne boosts and local Galilean transformations.
The resulting L\'evy-Leblond operator includes non-minimal couplings to the Newton-Cartan structure as well as to the gauge field,
and with these couplings it transforms covariantly.
Phenomenologically, this fixes the gyromagnetic ratio to $g\,{=}\,1$.
Three-dimensional spacetimes are an exception: generic $g$ is possible but results in modified Milne transformations,
which -- upon gauge fixing -- reproduces the anomalous diffeomorphisms found in earlier approaches.

\end{abstract}

\maketitle
\tableofcontents

\section{Introduction}
In recent years, Galilean invariant (quantum) field theories have found application in prominent
condensed matter systems like the unitary Fermi gas and (fractional) quantum hall states \cite{Son:2005rv,Hoyos:2011ez,2013arXiv1306.0638T,Gromov:2014vla}.
It is desirable to have a formulation of the theories on generic and possibly torsionful curved backgrounds,
both to incorporate the effects of e.g.\ non-trivial lattice backgrounds for condensed matter applications  \cite{Hughes:2012vg},
and to couple to gravity.
Newton-Cartan spacetimes are the non-relativistic analog of generic (pseudo-)Riemann manifolds.
They consist of a topological manifold $\mathcal M$ equipped with a nowhere vanishing one-form $n$ singling out a time direction
and a co-rank~$1$ symmetric tensor field $h^{-1}$, which gives an (inverse) spatial metric on constant-time slices.
We use $h^{-1}$ to indicate that it maps from the cotangent space to the tangent space,
the opposite direction compared to the metric in a pseudo-Riemannian structure,
but note that $h^{-1}$ is not defined as the inverse of some $h$.
The investigation of Newton-Cartan structures and field theories coupled to said structures has a long history
in the context of gravitational physics \cite{Künzle1972,Duval:1983pb,Duval:1984cj}, and has more recently
also found its way into holographic studies \cite{Christensen:2013lma,Christensen:2013rfa}.

In this work we focus on spinor fields, which have been considered previously on Newtonian spacetimes \cite{Kunzle:1984dt}.
Newtonian spacetimes are a special class of Newton-Cartan spacetimes where the clock form $n$ is required to be closed, $dn = 0$.
To our knowledge, an action for non-relativistic spinor fields in generic Newton-Cartan backgrounds
has not been constructed yet.
Considering the prevalence of spinor particles in condensed matter systems, we find the pursuit of such a construction
to be a worthwhile endeavor.
In addition to the motivations mentioned above, $n$ also needs to be left unconstrained in order to understand it as source for
the energy current, as emphasized in \cite{2015PhRvL.114a6802G,2015PhRvB..91l5303B}.
The starting point for our construction is the relativistic Dirac Lagrangian, from which we construct
a non-relativistic L\'evy-Leblond type action \cite{LevyLeblond:1967zz} by a limiting procedure, which
amounts to sending the speed of light to infinity and compensating the rest energy by a fine-tuned chemical potential.
This allows us to obtain completely general and even acausal Newton-Cartan backgrounds from relativistic parents.

Insisting on the existence of a non-trivial limit even in spacetimes with no standard notion of causality turns out to be remarkably fruitful.
We find that a non-minimal coupling to the background gauge field is needed, and that this non-minimal coupling is indeed
crucial even in the most benign backgrounds, to ensure invariance of the final Lagrangian under Milne boosts.
The final non-relativistic Lagrangian has all desired symmetries:
in addition to the symmetries of the Newton-Cartan structure,
it has local Galilean invariance reflecting the gauge redundancy in choosing a local frame as well as invariance under Milne boosts,
which reflects the gauge redundancy in choosing a dual vector field $v$ corresponding to the one-form $n$.\footnote{%
These invariances may be looked at either as spurionic symmetries, or,
if the theory is coupled to gravity, as gauge symmetries.
Either way, the actual global symmetries of the theory in a given background arise as the subset which leaves that background invariant.
For our purposes the two points of view are equivalent.
}
The non-minimal coupling that is required to guarantee a non-trivial limit and the presence of these symmetries results in a special value
for the gyromagnetic ratio of the non-relativistic spinor field, even on flat space.
This special value is $g\,{=}\,1$, half of the relativistic value.
Three-dimensional spacetimes are an exception and we discuss these separately, generalizing previous work in \cite{2015PhRvD..91d5030G}.

The paper is organized as follows.
In sec.~\ref{sec:newton-cartan-data} we set up notation for the relativistic parent geometry and introduce
the data that feeds into the non-relativistic limit.
We then turn to the actual limiting procedure in sec.~\ref{sec:NR-limit-general}, which contains as main result the non-relativistic Dirac
or generalized L\'evy-Leblond operator and Lagrangian in eqns. (\ref{eqn:massive-NR-Dirac})
and (\ref{eqn:NR-Lagrangian}), respectively.
In sec.~\ref{sec:symmetries} we discuss in detail the invariances that the relativistic symmetries induce in the non-relativistic limit.
In particular, we show that the generalized L\'evy-Leblond operator defined in eqn.\ (\ref{eqn:massive-NR-Dirac}) transforms covariantly under
Milne boosts and local Galilean transformations.
To have all mentioned symmetries manifest, we need the non-relativistic spinor to have as many components as the relativistic Dirac spinor has.
However, in the non-relativistic limit half of these components are downgraded to auxiliary fields.
In sec.~\ref{sec:causal-spacetimes} we solve for these auxiliary fields in causal spacetimes and give a Lagrangian purely in terms of the dynamical fields.
We also discuss explicitly the implications for flat space.
In sec.~\ref{sec:3d} we specialize to three-dimensional spacetimes, before closing with a discussion in sec.~\ref{sec:discussion}.

\section{Relativistic geometry and Newton-Cartan data}\label{sec:newton-cartan-data}
We start from a pseudo-Riemann manifold $(\mathcal M,g)$, and
for the non-relativistic limit we pick a nowhere-vanishing one-form $n$.
For each $p\in \mathcal M$, $n$ gives a preferred subspace of $T_p\mathcal M$: the ``spacelike'' vectors which are annihilated by $n$.
If $n\wedge dn=0$, the Frobenius theorem guarantees that there is a foliation of $\mathcal M$ by codimension-$1$ surfaces,
for which the tangent spaces consist exactly of said spacelike vectors (see e.g.\ the formulation in \cite{Wald:1984rg}).
This gives a foliation by spacelike hypersurfaces and a notion of causality.
However, we will not assume that $n\wedge dn=0$, but only that $n$ is timelike everywhere.
This ensures that the resulting $h^{\mu\nu}$ is positive definite,
as required for the Newton-Cartan structure.

With the clock form $n$ in hand, we can define an inverse spatial metric $h^{\mu\nu}$ from the inverse metric $g^{\mu\nu}$.
Any one-form $\lambda$ can be decomposed into spatial and timelike parts as
$\lambda=\lambda_s+n g^{-1}(\lambda,n)/g^{-1}(n,n)$, such that $\lambda_s$ is spatial, $g^{-1}(\lambda_s,n)=0$.%
\footnote{%
Spelling out the notation more explicitly, $g^{-1}(a,b)$ and $g(a,b)$ mean $g^{\mu\nu}a_\mu b_\nu$ and $g_{\mu\nu}a^\mu b^\nu$, respectively.
The one-forms are understood as $n=n_\mu dx^\mu$ and analogously for $\lambda$, $\lambda_s$.
}
The inverse spatial metric is then defined by $h^{-1}(\lambda,\tau)=g^{-1}(\lambda_s,\tau_s)$.
The rank of $h^{\mu\nu}$ is $d\,{-}\,1$ and it satisfies $h^{\mu\nu}n_\nu=0$.
The triple $(\mathcal M,n,h^{-1})$ is called a Galilean structure in \cite{Jensen:2014aia} and a Leibnizian structure in \cite{Bekaert:2014bwa}.

Since we start from a relativistic theory on a manifold with a pseudo-Riemannian structure,
we can immediately define a vector field $v$ to get a similar split of the tangent space, along with a spatial metric.
Since $g$ provides us with an isomorphism of $T_p\mathcal M$ and $T_p^\star M$,
we can simply define $v$ by $v^\mu=g^{\mu\nu}n_\nu/g^{-1}(n,n)$, s.t.\ $n_\mu v^\mu=1$.
We can then define projectors on the spatial part for tensors as
\begin{align}\label{eqn:tensor-projectors}
 \Pg\mu\nu&=g_\mu^{\hphantom{\mu}\nu}-n_\mu v^\nu~.
\end{align}
For the (inverse) spatial metrics we use the symbols $h^{-1}=h^{\mu\nu}\partial_\mu\otimes\partial_\nu$
and $h=h_{\mu\nu}dx^\mu\otimes dx^\nu$, with the understanding that $h^{\mu\nu}$ and $h_{\mu\nu}$ as matrices are not inverse to each other.
They are inverse as maps only when restricted to the spatial subspace singled out by $P_\mu^{\hphantom{\mu}\nu}$.
With the projectors (\ref{eqn:tensor-projectors}) we then have
\begin{align}
 h_{\mu\nu}&=\Pg\mu\rho\Pg\nu\sigma g_{\rho\sigma}~,&
 h^{\mu\nu}&=\Pg\rho\mu\Pg\sigma\nu g^{\rho\sigma}~.
\end{align}
For such a generic choice of $n$, the metric can then be decomposed in an ADM-like fashion as
\begin{align}\label{eqn:metric-split}
 g_{\mu\nu}&=\frac{n_\mu n_\nu}{g^{-1}(n,n)} +h_{\mu\nu}~,&
 g^{\mu\nu}&=g^{-1}(n,n)v^\mu v^\nu  +h^{\mu\nu}~.&
\end{align}
We fix $g^{-1}(n,n)=-c^{-2}$ in the following,
such that the split of the (inverse) metric reduces to the choices made in \cite{Jensen:2014wha},
namely $g_{\mu\nu}=-c^2n_\mu n_\nu+h_{\mu\nu}$ and $g^{\mu\nu}=-c^{-2}v^\mu v^\nu+h^{\mu\nu}$.
This does not restrict the class of Newton-Cartan geometries that can be obtained from the
reduction -- for given Newton-Cartan data one can always choose a suitable pseudo-Riemann metric such that our conditions are met.

From the Newton-Cartan perspective, the logic is quite disparate.
Without a pseudo-Riemann metric, $n$ and $h^{-1}$ alone do not determine a unique $v$ or $h$.
The conditions $n_\mu v^\mu=1$ and $v^\mu h_{\mu\nu}$ are invariant under Milne boosts,
which change $v$ and $h$ but leave $n$ and $h^{-1}$ invariant.
In the non-relativistic limit, this freedom in the choice of $v$ and $h$ arises as follows \cite{Jensen:2014wha}.
In the $c\rightarrow\infty$ limit, $n$ and $h^{-1}$ are the dominant structures in $g$ and $g^{-1}$, respectively.
That is,
\begin{align}
 c^{-2}g_{\mu\nu}\vert_{c\rightarrow\infty}&=-n_\mu n_\nu~,&
 g^{\mu\nu}\vert_{c\rightarrow\infty}&=h^{\mu\nu}~.
\end{align}
We can then extract $n$ from $c^{-2}g\vert_{c\rightarrow\infty}$ -- up to a sign ambiguity in ``taking the square root'' --
and $h^{-1}$ from $g^{-1}\vert_{c\rightarrow\infty}$.
The pseudo-Riemannian structure thus reduces to a Newton-Cartan structure when $c\rightarrow\infty$.
The derived quantities $v$ and $h$, on the other hand, only enter as subleading terms in $g^{-1}$ and $g$, respectively.
Redefining $n$ and $h^{-1}$ by subleading terms gives the same Newton-Cartan structure but changes $v$ and $h$, and
we indeed recover the Milne transformations.
We will see this more explicitly in sec.~\ref{sec:symmetries-milne}.

\subsection{Frames and spinors}
Now that we have discussed how the metric data is decomposed for a given $n$, we turn to the objects associated with the frame bundle:
the vielbein (solder form), spin connection and spinors.
From the decomposition of $g_{\mu\nu}$ in terms of $n_\mu$, $h_{\mu\nu}$, we immediately get an induced decomposition of the Minkowski metric
in terms of $n_a=e_a^\mu n_\mu$, namely
\begin{align}
 \eta_{ab}&=-c^2 n_a n_b+ h_{ab}~,
 &h_{ab}&=\Pg{a}c\Pg{b}d \eta_{cd}~,
 &P^a_{\hphantom{a}b}=\eta^a_{\hphantom{a}b}-v^a n_b~.
\end{align}
The decomposition of $g^{\mu\nu}$ in terms of $v^\mu$ and $h^{\mu\nu}$ likewise induces a decomposition of $\eta^{ab}$.
We also have $h_{\mu\nu}=e_\mu^a e_\nu^b h_{ab}$ and analogously for $h^{\mu\nu}$.
The (inverse) vielbein can correspondingly be decomposed as
\begin{align}
 e_\mu^a&=n_\mu v^a + \tilde e_\mu^a~,
 &e^\mu_a&=v^\mu n_a + \tilde e^\mu_a~,
\end{align}
where the tilde denotes a spatial projection on the Lorentz index,
$\tilde e_\mu^a=\Pg{b}a e_\mu^b$ and $\tilde e^\mu_a=\Pg{a}b e^\mu_b$.
Occasionally we will find it useful to explicitly split Lorentz indices as $a=(\underline{0},i)$,
and we generally use an underline to distinguish Lorentz from coordinate indices whenever explicit values are used.
With the Minkowski metric $\eta=-c^2dt^2+d\vec{x}^2$ and the two decompositions of $g$ as
\begin{align}
 g_{\mu\nu}&=-c^2n_\mu n_\nu+h_{\mu\nu}~,
 &g_{\mu\nu}&=-c^2 e^{\underline{0}}_\mu e^{\underline{0}}_\nu +e^{i}_\mu e^{j}_\nu \delta_{ij}~,
\end{align}
we immediately find
\begin{align}\label{eqn:e0-decomposition}
 e^{\underline{0}}&=n+\frac{1}{c^2} a~,
\end{align}
where $a$ is $\mathcal O(1)$.
To keep the notation clean we will partly use index-free notation to suppress coordinate indices,
e.g.\ for the above $e^{\underline{0}}=e^{\underline{0}}_\mu dx^\mu$, $n=n_\mu dx^\mu$ and $a=a_\mu dx^\mu$ are understood -- they are all one-forms.
We also note that $n_b=\delta_{b\underline{0}}-c^{-2}a_b$,
which will be useful below.
We then have $e^{\underline{0}}\vert_{c\rightarrow\infty}=n$, which makes $e^a\vert_{c\rightarrow\infty}$
into a Galilei frame as defined in \cite{Künzle1972}.
The subleading part of the vielbein, $a$, plays the role of the mass gauge field in \cite{Geracie:2015dea}.
Note that, since $e^{\underline{0}}$ transforms under local Lorentz transformations while $n$ does not,
$a$ transforms non-trivially under local Lorentz boosts, in spite of not having a Lorentz index.
We will see in sec.~\ref{sec:symmetries} that the orders in $c$ work out.

We now turn to the spin connection.
In the relativistic theory we start with a torsion-free background\footnote{%
When coupled to gravity in a $1^\mathrm{st}$-order formalism, spinor fields produce torsion
when the spin connection is determined from its equation of motion.
That torsion is quadratic in the Dirac field and would show up in the Lagrangian.
However, it drops out once the gravitational coupling is sent to zero.
One can nevertheless study torsionful backgrounds as a starting point, which we leave for the future.
},
so the spin connection
can be expressed entirely in terms of the vielbein as
\begin{align}\label{eqn:relativistic-spin-connection}
 \omega_{\mu\nu\rho}&=3\Omega_{[\mu\nu\rho]}-2\Omega_{\nu\rho\mu}~,
 &\Omega_{\mu\nu\rho}=e_{\rho a}\partial_{[\mu}e_{\nu]}^a~.
\end{align}
Converting all indices to coordinate indices obscures the structure of $\omega$ as an $\mathfrak{so}(1,d-1)$-valued one form,
but we find it useful for explicit computations.
Despite the absence of Lorentz indices, this expression for the spin connection still transforms non-trivially under local Lorentz transformations,
as appropriate for a connection.
We isolate the divergent pieces in $\omega$ as follows.
With the vielbein as discussed above, we have
$\Omega_{\mu\nu\rho}=-c^2 e_{\rho}^{\underline{0}}\partial_{[\mu}e_{\nu]}^{\underline{0}}+\delta_{ij}e_\rho^i\partial_{[\mu}e_{\nu]}^j$.
Using eqn.\ (\ref{eqn:e0-decomposition}) we then find
\begin{align}\label{eqn:spin-connection-expansion}
 \omega_{\mu\nu\rho}&=c^2 \left(2n_\mu \partial_{[\nu}n_{\rho]}-3 n_{[\mu}\partial_\nu n_{\rho]}\right)
 +\dot\omega_{\mu\nu\rho}~,
\end{align}
where $\dot\omega_{\mu\nu\rho}$ is $\mathcal O(1)$.
Explicitly, we have
\begin{align}\label{eqn:spin-connection-etilde}
 \dot\omega_{\mu\nu\rho}&=3\dot\Omega_{[\mu\nu\rho]}-2\dot\Omega_{\nu\rho\mu}~,
 &\dot\Omega_{\mu\nu\rho}&= e_{\rho i}\partial_{[\mu} e_{\nu]}^i
 -a_\rho\partial_{[\mu}n_{\nu]}
 -n_\rho\partial_{[\mu}a_{\nu]}
 -c^{-2}a_\rho\partial_{[\mu}a_{\nu]}~.
\end{align}

Finally, we come to the spinors themselves and to the Clifford algebra.
We decompose the generators as
\begin{align}
 \gamma^\mu&=\slashed{n}v^\mu+\tilde\gamma^\mu~,
 &\slashed{n}&=n_\mu\gamma^\mu~,
 &\tilde\gamma^\mu&=\Pg\nu\mu\gamma^\nu~,
\end{align}
and analogously for lowered spacetime and Lorentz indices.
With $n$ as additional geometric data, we can
define a new ``chirality'' operator for the spinors, similarly to the tensor projectors in (\ref{eqn:tensor-projectors}).
This is regardless of whether the relativistic spin group admits chiral representations.
Namely,
\begin{align}\label{eq:spinor-projectors}
 P_\pm&=\frac{1}{2}(\mathds{1}\pm i c \slashed{n})~.
\end{align}
In the non-relativistic limit we expect a pair of spinors with half as many components
as the relativistic Dirac spinor, which differ in their dynamics \cite{LevyLeblond:1967zz}.
It is thus natural to expect projectors to be relevant for the reduction.
With $P_\pm$ we decompose the Dirac field $\psi$ as
\begin{align}
 \psi&=\psi_++\psi_-~,&\psi_\pm&=P_\pm\psi~.
\end{align}
With $\gamma^{\underline{0}}\gamma_\mu^\dagger\gamma_{\underline{0}}=-\gamma_\mu$ we have
$\overline{P_\pm}=\gamma^{\underline{0}}(P_\pm)^\dagger\gamma_{\underline{0}}=P_\pm$
and thus $\overline{P_\pm\psi}=\overline{\psi}P_\pm$.

\section{Non-relativistic limits in generic backgrounds}\label{sec:NR-limit-general}
We now set up and perform the non-relativistic limit for a Dirac field in generic backgrounds.
We start with a free Dirac field coupled non-minimally to a background gauge field $C$, which sources
the U(1) particle number current.
The Lagrangian is
\begin{align}\label{eqn:non-minimally-coupled-Lagrangian}
 L&=i\overline{\psi}\slashed{D}\psi-i\left(D_\mu\overline{\psi}\right)\gamma^\mu\psi - 2 i m c\overline{\psi}\psi
    +\frac{\alpha}{mc}F_{\mu\nu}\overline{\psi}\gamma^{\mu\nu}\psi~,
\end{align}
where $F=dC$ or $F_{\mu\nu}=\partial_\mu C_\nu-\partial_\nu C_\mu$.
The reason for including the non-minimal coupling will become apparent below.
The covariant derivative reads
\begin{align}\label{eqn:Dmu-def}
 D_\mu\psi=\partial_\mu\psi+\frac{1}{4}\omega_{\mu ab}\gamma^{ab}-i C_\mu~.
\end{align}
Our convention for the Clifford algebra generators is $\lbrace \gamma^a,\gamma^b\rbrace=2\eta^{ab}\mathds{1}$
with $\gamma^{\underline{0}}\gamma_a^\dagger \gamma_{\underline{0}}=-\gamma_a$,
and we use $\overline{\psi}=\psi^\dagger c\gamma^{\underline{0}}$ to have $\overline \psi$ of the same order in $c$ as $\psi$.
The action of $D_\mu$ on $\overline{\psi}$ is determined from $\overline{\psi}\psi$ being a scalar and the Leibniz rule.
Moreover, we have $\omega$ and $C$ real and $i D_\mu$ is formally self-adjoint with respect to the usual inner product for spinors.
To set the stage, we determine the scaling weights of $\psi_\pm$ as $c\rightarrow\infty$ from a flat-space analysis.
Our guideline will be to obtain non-trivial dynamics at $c\rightarrow\infty$.
The reduction in generic backgrounds will be done afterwards.

\subsection{Scaling weights from flat space}
We take Minkowski space with $g=-c^2dt^2+d\vec{x}^2$ and fix $e_\mu^a=\delta_\mu^a$.
For the non-relativistic limit we choose $n=dt$.
We note that fixing an expression for $n$ to all orders in $c$ also fixes a Milne frame, and we have $v=\partial_t$.
The Lagrangian given by eqn.\ (\ref{eqn:non-minimally-coupled-Lagrangian}) becomes
\begin{align}\label{eqn:Lagrangian-flat}
\begin{split}
 L&=i\overline{\psi}_\pm \tilde\gamma^\mu\left(\partial_\mu-i C_\mu\right)\psi_\mp
 \pm\frac{1}{c}\overline{\psi}_\pm\big(\partial_0-i C_0\big)\psi_\pm
 -imc\overline{\psi}_\pm\psi_\pm
 \\&\hphantom{=}\,
 +\frac{\alpha}{2mc}F_{\mu\nu}\overline{\psi}_\pm\tilde\gamma^{\mu\nu}\psi_\pm
 \pm\frac{i\alpha}{mc^2}F_{\mu\nu}v^\nu\overline{\psi}_\pm\tilde\gamma^\mu \psi_\mp
 +\mathrm{c.c.}~,
\end{split}
\end{align}
where a sum over the upper and lower choices for the subscripts is implicit.
The dominant part of the mass term will dominate all derivative terms at $c\rightarrow\infty$,
and it therefore needs to be canceled in order to get a non-trivial non-relativistic theory.
The lesson from \cite{Jensen:2014wha} is to turn on an $\mathcal O(c^2)$ background gauge field
giving a chemical potential as
\begin{align}\label{eqn:gauge-field-expansion}
 C_\mu &=-mc^2 n_\mu +mA_\mu~,
\end{align}
where the $\mathcal O(1)$ part, $A$, becomes the non-relativistic gauge field.
We will see below that this is also the correct choice for curved spaces, unless
the number of spacetime dimensions is three, a case we will discuss separately.
For the scaling analysis we set $A=0$, and noting that $dn=0$ the Lagrangian (\ref{eqn:Lagrangian-flat}) with the gauge field
(\ref{eqn:gauge-field-expansion}) then becomes
\begin{align}\label{eqn:Lagrangian-flat2}
 L&=i\overline{\psi}_\pm \tilde\gamma^\mu\partial_\mu\psi_\mp
 \pm\frac{1}{c}\overline{\psi}_\pm\partial_0\psi_\pm
 \pm imc\overline{\psi}_\pm\psi_\pm
 -imc\overline{\psi}_\pm\psi_\pm
 +\mathrm{c.c.}~,
\end{align}
The mass-like term resulting from the leading piece of eqn.\ (\ref{eqn:gauge-field-expansion})
has opposite signs for the two chiralities, allowing for a cancellation of only one of the actual mass terms.
With our choice of $C$ this is the one for $\psi_+$.
For $\psi_-$, the mass term dominates its time-derivative piece,
and we get a non-dynamical auxiliary field, as expected from \cite{LevyLeblond:1967zz,Kunzle:1984dt}.

For generic scaling of $\psi_\pm$, the equations of motion for $\psi_\pm$ decouple at $c\rightarrow\infty$.
Depending on which of the terms is dominant,
the equation resulting from varying $\psi_-$ then is either $\psi_-=0$ or $\tilde\gamma^\mu\partial_\mu\psi_+=0$.
Neither of those yields non-trivial dynamics for $\psi_+$ when combined with the equation resulting from variation of $\psi_+$.
To get a non-trivial limit, the mixed term in (\ref{eqn:Lagrangian-flat2}) has to contribute at the same order as the mass term for $\psi_-$.
That is, $\psi_+=\mathcal O(c\psi_-)$.
The overall scaling of the Lagrangian can be absorbed by a rescaling at the end, so we simply fix $\psi_\pm=\mathcal O(c^{\pm 1/2})$.

\subsection{Non-relativistic limit in generic backgrounds}
We now perform the non-relativistic limit for generic curved backgrounds and generic choice of $n$.
From \cite{LevyLeblond:1967zz} we expect that we need to keep all components of the relativistic Dirac spinor.
With the scaling determined from the flat case, we define our non-relativistic spinor, $\Psi$, as
\begin{align}\label{eqn:NR-spinor-def}
 \Psi&=\mathcal P\psi\big\vert_{c\rightarrow\infty}~,& \mathcal P&=c^{-\frac{1}{2}}P_+ + c^{\frac{1}{2}}P_-~.
\end{align}
Note that $\Psi$ is $\mathcal O(1)$ and indeed keeps as many non-trivial components as the Dirac spinor we started with.
We will justify calling it a non-relativistic spinor in sec.~\ref{sec:symmetries}.

We will disassemble the Lagrangian into its building blocks as far as possible.
To perform the non-relativistic limit we write the Lagrangian of eqn.\ (\ref{eqn:non-minimally-coupled-Lagrangian}) as
\begin{align}\label{eqn:non-minimally-coupled-Lagrangian-2}
 L&=i\overline{\psi}\left(\slashed{D}-mc -\frac{i\alpha}{2mc}\slashed{F}\right)\psi+\cc~,&
 \slashed{F}&=F_{\mu\nu}\gamma^{\mu\nu}~.
\end{align}
The operator $\mathcal P$ is invertible, and we define $\Pt$ such that $\Pt \mathcal P=\mathcal P \Pt=\mathds{1}$.
This allows us to conveniently insert identities in the Lagrangian (\ref{eqn:non-minimally-coupled-Lagrangian-2}),
and with the definition of $\Psi$ in eqn.\ (\ref{eqn:NR-spinor-def}) we see
\begin{align}\label{eqn:Ptilde-def}
 L&=i\overline{\Psi}\,\Pt\left(\slashed{D}-mc -\frac{i\alpha}{2mc}\slashed{F}\right)\Pt\,\Psi+\cc~,
 &\Pt&=c^\frac{1}{2}P_+ + c^{-\frac{1}{2}}P_-~.
\end{align}
We want to define the non-relativistic Lagrangian as $L_\mathrm{nr}=L\vert_{c\rightarrow\infty}$,
and to obtain a non-trivial limit we will have to fix $\alpha$.
Since $\Psi$ is $\mathcal O(1)$ by definition, it just remains to evaluate what
will become the definition of our non-relativistic Dirac operator.
Namely,
\begin{align}\label{eqn:Dnr-def}
 \Dnr_\alpha&:=\Pt\left(\slashed{D}-mc -\frac{i\alpha}{2mc}\slashed{F}\right)\Pt\Big\vert_{c\rightarrow\infty}~.
\end{align}
The covariant derivative was defined in eqn.\ (\ref{eqn:Dmu-def}), and we choose $C$ in accordance with eqn.\ (\ref{eqn:gauge-field-expansion}).
We will also need the spin connection as evaluated in eqn.\ (\ref{eqn:spin-connection-expansion}).
With these results, we find for the partial derivative and gauge field parts of $\Dnr_\alpha$
\begin{align}\label{eqn:Dnr-evaluation-1}
 \Pt \gamma^\mu(\partial_\mu-i C_\mu)\Pt\Big\vert_{c\rightarrow\infty}&=
 \left(-i P_+v^\mu+\tilde\gamma^\mu\right)\Big(\partial_\mu-\frac{i}{2}(\partial_\mu a_b)e_\nu^b \tilde \gamma^\nu P_+-i m A_\mu\Big)
 +mc^2P_+ - m P_-~.
\end{align}
For quantities which are $\mathcal O(1)$ at $c\rightarrow\infty$, like $e^a$ and $A$, we use the same symbol to denote the
non-relativistic object, with the understanding that the latter is the leading part only.
This applies to the right hand side of the above equation and to avoid unnecessarily complicated notation we will use that
convention in the following when $c\rightarrow\infty$ is clear from the context.
The mass term is simply $\Pt mc\Pt=mc^2P_+ +m P_-$, and we see the divergent piece here cancel against that of eqn.\ (\ref{eqn:Dnr-evaluation-1}).
The remaining terms of $\Dnr_\alpha$ contain the spin connection as well as the non-minimal coupling.
As seen from eqn.\ (\ref{eqn:spin-connection-expansion}), the spin connection has an $\mathcal O(c^2)$ piece.
Looking, then, at the divergent part of the spin connection and non-minimal coupling pieces of $\Dnr_\alpha$, we find
\begin{subequations}\label{eqn:divergence-Omega-F}
\begin{align}
 \Pt\left(\Omega-\frac{i\alpha}{2mc}\slashed{F}\right)\Pt\Big\vert_{\mathcal O(c^2)}&=
 \frac{c^2}{8} \partial_{[\mu}n_{\nu]} P_+ \left(c\slashed{n}\gamma^{\mu\nu}-3\gamma^{\mu\nu}c\slashed{n}+8i\alpha\gamma^{\mu\nu}\right)P_+
 \\
 &=ic^2\left(\frac{1}{4}+\alpha\right)P_+\partial_{[\mu} n_{\nu]}\gamma^{\mu\nu}P_+~,
\end{align}
\end{subequations}
where $\Omega=\frac{1}{4}\gamma^\mu\omega_{\mu ab}\gamma^{ab}$.
If $\alpha\neq -\frac{1}{4}$, this term dominates the Lagrangian at large $c$.
One could imagine modifying the leading behavior of $C$ to cancel this divergence, e.g.\
by adding a term proportional to $dn$ to eqn.\,(\ref{eqn:gauge-field-expansion}).
This will indeed work for three-dimensional spacetimes, which we discuss in sec.~\ref{sec:3d}.
Generically, however, the divergence can not be canceled this way due to the different Clifford algebra structures of $\Omega$ and $\slashed{C}$.
A possible workaround one may think of is modifying the projectors $P_\pm$
by adding terms like $(n\wedge dn)_{\mu\nu\rho}\gamma^{\mu\nu\rho}$,
but ensuring that the projectors square to themselves then is problematic
and we did not find a reasonable solution from this route.
A different variant of the non-minimal coupling term to cancel the divergence
would be $\overline{\psi}(C\wedge dC)_{\mu\nu\rho}\gamma^{\mu\nu\rho}\psi$, but this is not gauge invariant.
We are thus lead to conclude that in order to find non-trivial dynamics, we must fix
\begin{align}\label{eqn:alpha}
 \alpha=-\frac{1}{4}~.
\end{align}
Before moving on, we want to elaborate a bit on the divergent terms in eqn.\ (\ref{eqn:divergence-Omega-F}).
If $n\wedge dn=0$, we can use $\lbrace\gamma^\mu,\gamma^{\nu\rho}\rbrace=2\gamma^{\mu\nu\rho}$ to see that
\begin{align}\label{eqn:zero-ndn-1}
 P_\pm \gamma^{\nu\rho}\partial_{[\nu} n_{\rho]}P_\pm&=
 \pm ic P_\pm n_\mu\lbrace\gamma^\mu,\gamma^{\nu\rho}\rbrace P_\pm \partial_{[\nu} n_{\rho]}
 =\pm 2icP_\pm \gamma^{\mu\nu\rho} P_\pm n_{[\mu}\partial_{\nu} n_{\rho]}
 =0~.
\end{align}
That means as long as $n\wedge dn=0$, the divergent terms in eqn.\ (\ref{eqn:divergence-Omega-F}) actually vanish by themselves,
and one might hope to get away without fixing $\alpha=-\frac{1}{4}$.
We will see below that symmetry considerations nevertheless call for the choice of $\alpha$ in eqn.\ (\ref{eqn:alpha}), even on flat space.
The reason is simple: while the divergent terms themselves vanish for $n\wedge dn=0$, their Milne variations for generic $\alpha$
propagate into the $\mathcal O(1)$ part, such that the final expression for $\Dnr$ in eqn.\ (\ref{eqn:massive-NR-Dirac}) does not transform covariantly.
We come back to Milne variations more explicitly in sec.~\ref{sec:symmetries-milne}.
With the choice of $\alpha = -\frac{1}{4}$, $\Pt\left(\Omega-\frac{i\alpha}{2mc}\slashed{F}\right)\Pt$ is indeed finite
for generic Newton-Cartan backgrounds, and we find
\begin{align}\label{eqn:Dnr-evaluation-2}
\begin{split}
  \Pt\left(\Omega+\frac{i}{8mc}\slashed{F}\right)\Pt\Big\vert_{c\rightarrow\infty}=\,&
  \left(-i P_+v^\mu+\tilde\gamma^\mu\right)\frac{1}{4}\dot\omega_{\mu\nu\rho}\left(\tilde\gamma^{\nu\rho}-2i\tilde\gamma^\nu v^\rho P_+\right)
  +\frac{i}{8} F_{\mu\nu}^{(A)}\tilde\gamma^{\mu\nu} P_+
  \\
  &+\frac{1}{2}T_{\mu\nu}^c v^\nu P_+ \tilde\gamma^\mu
   -\frac{i}{4}T_{\mu\nu}^\mathrm{c}\tilde\gamma^{\mu\nu}P_-
   +\frac{1}{4}T_{\mu\nu}^c \tilde\gamma^\mu v^\nu~,
\end{split}
\end{align}
where $F^{(A)}=dA$ and $T^\mathrm{c}=dn$.
For $\Dnr_{-1/4}$ we then find from eqns. (\ref{eqn:Dnr-evaluation-1}), (\ref{eqn:Dnr-evaluation-2}) and the mass term that
\begin{align}\label{eqn:Dnr-explicit}
 \Dnr_{-\frac{1}{4}}=\,&
 \left(-i P_+v^\mu+\tilde\gamma^\mu\right)\mathcal D_\mu
 -2m P_- +\frac{i}{8} F^{(A)}_{\mu\nu}\tilde\gamma^{\mu\nu}P_+
 -\frac{i}{4}T_{\mu\nu}^\mathrm{c}
 \left(\tilde\gamma^{\mu\nu}P_- -2iv^\mu P_+\tilde\gamma^\nu+i\tilde\gamma^\mu v^\nu\right)~.
\end{align}
Note that the $P_\pm$ projectors when acting on $\Psi$ are $\mathcal O(1)$.
The covariant derivative $\mathcal D_\mu$ is defined as
\begin{align}\label{eqn:covariant-derivative}
 \mathcal D_\mu&:=\partial_\mu+\frac{1}{4}\hat\omega_{\mu\nu\rho}\left(\tilde\gamma^{\nu\rho}-2i\tilde\gamma^\nu v^\rho P_+\right)-imA_\mu~.
\end{align}
The final non-relativistic spin connection $\hat\omega$ used in (\ref{eqn:covariant-derivative}) is given by
$\hat\omega_{\mu\nu\rho}=\dot\omega_{\mu\nu\rho}+e_\nu^b n_\rho\partial_\mu a_b\vert_{c\rightarrow\infty}$.
More explicitly, with $f=da$ it is given by
\begin{align}\label{eqn:spin-connection-hat}
 \hat\omega_{\mu\nu\rho}&=3\hat\Omega_{[\mu\nu\rho]}-2\hat\Omega_{\nu\rho\mu}+e_\nu^b n_\rho\partial_\mu a_b~,
 &
 \hat\Omega_{\mu\nu\rho}&= e_{\rho i}\partial_{[\mu} e_{\nu]}^i
 -\frac{1}{2}a_\rho T_{\mu\nu}^\mathrm{c}
 -\frac{1}{2}n_\rho f_{\mu\nu}~.
\end{align}
The conversion to frame indices proceeds via 
$\hat\omega_{\mu\hphantom{a}b}^{\hphantom{\mu}a}=\eta^{ac}\hat\omega_{\mu\nu\rho}e^\nu_c e^\rho_b\vert_{c\rightarrow\infty}=
h^{ac}\hat\omega_{\mu\nu\rho}e^\nu_c e^\rho_b$.
%We will discuss the structure in sec.~\ref{sec:symmetries-Galilean}.
%
We take the expression for $\Dnr_{-1/4}$ in eqn.\ (\ref{eqn:Dnr-explicit}),
which is given purely in terms of Newton-Cartan data and has the divergent pieces canceled,
to define $\Dnr$ without a subscript, which finally is our non-relativistic Dirac or L\'evy-Leblond operator.
With $\dot\gamma^\mu=-iv^\mu P_+$ it reads
\begin{subequations}\label{eqn:massive-NR-Dirac}
\begin{align}
 \Dnr:=&\,
 \left(\dot\gamma^\mu +\tilde\gamma^\mu\right)\mathcal D_\mu
 -2m P_-
 +\frac{i}{8} F_{\mu\nu}^{(A)}\tilde\gamma^{\mu\nu}P_+
 -\frac{i}{4}T_{\mu\nu}^\mathrm{c} \left(\tilde\gamma^{\mu\nu}P_-+2\dot\gamma^\mu \tilde\gamma^\nu+i\tilde\gamma^\mu v^\nu\right)
 ~,
\\
\mathcal D_\mu=&\,\partial_\mu+\frac{1}{4}\hat\omega_{\mu\nu\rho}\left(\tilde\gamma^{\nu\rho}+2\tilde\gamma^\nu \dot\gamma^\rho\right)-imA_\mu~.
\label{eqn:calD-mu-def}
\end{align}
\end{subequations}
The operator $\mathcal D_\mu$ looks strikingly similar to the definition of the covariant derivative in \cite{Kunzle:1984dt}.
It is different in that our Clifford algebra is not directly the one associated to the degenerate
bilinear form given by $h^{\mu\nu}$:
our $\dot\gamma^a n_a$, which corresponds to $\gamma^0$ in \cite{Brooke:1978tr,Kunzle:1984dt}, squares to itself rather than to zero.
We will see below that the spin group constructed from our Clifford algebra nevertheless agrees with \cite{Brooke:1978tr,Kunzle:1984dt}.
With the manifestly finite (massive) non-relativistic Dirac operator (\ref{eqn:massive-NR-Dirac}),
we then find the general non-relativistic spinor Lagrangian
\begin{align}\label{eqn:NR-Lagrangian}
 L_\mathrm{nr}=i\overline{\Psi}\Dnr\Psi+\cc~.
\end{align}
This expression for the Lagrangian itself hides all the details,
like couplings to the clock torsion on spacetimes with $dn\neq 0$ etc.
These are manifest in the non-relativistic Dirac operator $\Dnr$ in (\ref{eqn:massive-NR-Dirac}).
We will see below that $\Dnr$, i.e.\ the complete operator, is the preferred one from the symmetry perspective.
Finally, we note that the volume form needed to construct an action from this Lagrangian is just the standard
Newton-Cartan volume form discussed e.g.\ in \cite{Jensen:2014aia}.
Namely, $\operatorname{vol}=\sqrt{\det\upgamma}\,dx^0\wedge\dots \wedge dx^{d-1}$, where $\upgamma_{\mu\nu}=n_\mu n_\nu+h_{\mu\nu}$.

\section{Symmetries}\label{sec:symmetries}
Our main focus will be on Milne and local Galilean transformations, which we discuss in detail in the next two subsections.
Under diffeomorphisms the vielbein and spin connection transform covariantly.
The spinor field transforms as a spinor under local Galilean transformations, but as a scalar under diffeomorphisms.
The Lagrangian (\ref{eqn:NR-Lagrangian}) thus transforms as a scalar and the fact that we are considering spinor fields
does not interfere with the symmetries of the Newton-Cartan structure.
The U(1) gauge symmetry also only needs a brief discussion.
The gauge field $A$ inherits the gauge transformations from those of $C$ via eqn.\ (\ref{eqn:gauge-field-expansion}), since $n$ is invariant.
The projector $\mathcal P$ is gauge invariant, so the transformation of $\Psi$ is the same as for $\psi$,
and since and the derivative in eqn.\ (\ref{eqn:calD-mu-def}) is gauge covariant, $\Dnr$ transforms correctly.
As a result, the Lagrangian in eqn.\ (\ref{eqn:NR-Lagrangian}) is invariant, as desired.

\subsection{Milne boosts}\label{sec:symmetries-milne}
As discussed in \cite{Jensen:2014wha}, shifting the clock form, $n$, by certain one-forms which are
subleading in $c$ does not alter the Newton-Cartan structure obtained in the $c\rightarrow\infty$ limit.
We discuss this systematically, adding some details to the analysis of \cite{Jensen:2014wha},
before turning to the invariance of our non-relativistic Lagrangian.

\subsubsection{Milne transformations}
Since $g_{\mu\nu}$ does not depend on the decomposition into $n$ and $h$,
we can immediately infer how $h_{\mu\nu}$ transforms under shifts in $n$. Namely, we let
\begin{align}\label{eqn:Milne-shift-n}
 n_\mu&\rightarrow n_\mu^\prime=n_\mu - c^{-2}\xi_\mu~,&
 h_{\mu\nu}&\rightarrow h^\prime_{\mu\nu}=h_{\mu\nu}-\xi_{\mu}n_{\nu}-\xi_{\nu}n_{\mu}+c^{-2}\xi_\mu\xi_\nu~.
\end{align}
In the relativistic theory we have $v^\mu=g^{\mu\nu}n_\nu/g^{-1}(n,n)$, as explained in sec.~\ref{sec:newton-cartan-data}.
Using that
$v^{\prime\mu}=g^{\mu\nu}n^\prime_\nu/g^{-1}(n^\prime,n^\prime)$
and that $g^{-1}$ is invariant, we find
\begin{align}\label{eqn:delta-M-v-0}
 v^\mu\rightarrow v^{\prime \mu}=\frac{v^\mu+h^{\mu\nu}\xi_\nu-c^{-2}v^\mu\,\xi(v)}{-c^2g^{-1}(n^\prime,n^\prime)}~.
\end{align}
Demanding that $h_{\mu\nu}$ remains of rank $d-1$ with null eigenvector $v^\prime$,
i.e.\ $h^\prime_{\mu\nu}v^{\prime\nu}=0$, implies
\begin{align}\label{eqn:milne-xi-condition}
 \xi_\mu h^{\mu\nu}\xi_\nu+2 \xi_\mu v^\mu&=c^{-2}(\xi_\mu v^\mu)^2~.
\end{align}
This same constraint also ensures that the norm of $n$ is invariant, $g^{-1}(n^\prime,n^\prime)=-c^{-2}$.
The constraint in eqn.\ (\ref{eqn:milne-xi-condition}) is quadratic in $\xi$, and we expect two branches of solutions.
To understand these solutions, we split $\xi$ into timelike and spatial parts
\begin{align}\label{eqn:xi-decomposition}
 \xi_\mu&=n_\mu \lambda + \zeta_\mu~,&
 v^\mu\zeta_\mu&=0~,&
 \zeta_\mu\zeta^\mu=(c^{-2}\lambda-2)\lambda~.
\end{align}
The last equation fixing $\lambda$ in terms of $\zeta$ implements the constraint (\ref{eqn:milne-xi-condition}),
and the indices of the purely spatial $\zeta_\mu$ can be freely raised and lowered with the spatial metric, i.e.\ $\zeta^\mu=h^{\mu\nu}\zeta_\nu$.
Note that $\lambda\rightarrow 2c^{2}-\lambda$ leaves the right hand side of the last equation in (\ref{eqn:xi-decomposition}) invariant.
We find two solutions for $\lambda$,
\begin{align}\label{eqn:lambda-sol}
 \lambda(\zeta)&=c^2-c\sqrt{c^2+\zeta_\mu\zeta^\mu}~,&
 \tilde\lambda(\zeta)&=2c^2-\lambda(\zeta)~.
\end{align}
The first solution is $\mathcal O(1)$ in the large-$c$ limit, and choosing
it gives a family of transformations which are connected to the identity:
as $\zeta^\mu\zeta_\mu\rightarrow 0$ also $\lambda(\zeta)\rightarrow 0$,
and thus $\xi\rightarrow 0$.
These are what we will call Milne boosts.
The transformations with the second choice are not connected to the identity:
for $\zeta^\mu\zeta_\mu\rightarrow 0$ we get $\tilde\lambda(\zeta)\rightarrow 2c^2$.
In contrast to the other choice, this does change the Newton-Cartan data.
It corresponds to $n\rightarrow -n$ while $h_{\mu\nu}$ stays the same, so this is similar to a time reflection.
This transformation is compatible with the condition on the norm of $n$, and with $h^{-1}$ remaining corank $1$.
The pseudo-Riemann metric $g$ does not need to have any discrete or continuous isometries for this second branch to exist.
Note, however, that this is different from the notion of time reflection symmetry in the relativistic theory,
and this ambiguity in ``taking the square root'' has been well appreciated in, e.g., \cite{Künzle1972,Kunzle:1984dt,Duval:1984cj}.
The entire second branch can be obtained by combining the ``time reflection'' with the Milne boosts connected to the identity.

With the constraint (\ref{eqn:milne-xi-condition}) and either choice of solution for $\lambda$, the expression for $v^\prime$ in eqn.\ (\ref{eqn:delta-M-v-0})
simply evaluates to
\begin{align}
 v^{\prime \mu}&=v^\mu\left(1-c^{-2}\lambda\right)+\zeta^\mu~.
\end{align}
This turns into $v\rightarrow -v$ for the time reflection, as expected.
For the inverse spatial metric we find
\begin{align}\label{eqn:h-inv-Milne-transf}
 h^{\prime\mu\nu}&=h^{\mu\nu}+c^{-2}\left(\zeta^\mu\zeta^\nu+2(1-c^{-2}\lambda)v^{(\mu}\zeta^{\nu)}
  \right)+c^{-4}v^\mu v^\nu \zeta^\rho\zeta_\rho~.
\end{align}
It satisfies $n^\prime_\mu h^{\mu\nu}$ and is invariant under time reflection, again as expected.
The vielbein does not change under Milne boosts, but the decomposition of $e^{\underline{0}}$ which
we set up in eqn.\ (\ref{eqn:e0-decomposition}) does.
Similarly, the relativistic gauge field $C$ is also invariant, but since the non-relativistic gauge field $A$ arises from the decomposition of $C$,
via (\ref{eqn:gauge-field-expansion}), $A$ transforms as well.
We find
\begin{align}\label{eqn:aA-Milne-trafo}
 a&\rightarrow a+\xi~,
 &A&\rightarrow A-\xi~.
\end{align}

To complete the discussion, we give the Milne transformations of the non-relativistic quantities.
That is, the leading order in $c$ of the above discussion.
The quantities $n$, $h^{\mu\nu}$, $e_\mu^a$ and $e^\mu_a$ are invariant,
while
\begin{align}\label{eqn:Milne-leading-v-h}
 v^\mu&\rightarrow v^\mu+\zeta^\mu~,&
 h_{\mu\nu}&\rightarrow h_{\mu\nu}-2\zeta_{(\mu}n_{\nu)}+n_\mu n_\nu \zeta^2~,
\end{align}
along with $a\rightarrow a+\xi$ and $A\rightarrow A-\xi$,
where $\xi\vert_{c\rightarrow\infty}=\zeta-\frac{1}{2}n \zeta^2$.
Note that this transforms only objects associated directly with the (co)tangent bundle.
The vielbein, which solders the frame bundle to the tangent bundle, is invariant.

\subsubsection{Milne invariance of the non-relativistic Lagrangian}
\label{sec:milneinvofL}
We want to verify that the non-relativistic Lagrangian, eqn.\ (\ref{eqn:NR-Lagrangian}), is invariant under Milne boosts.
One might expect that any Lagrangian resulting from a non-relativistic limit should be Milne invariant automatically,
since the relativistic theory is.
That was part of the motivation to study non-relativistic limits in \cite{Jensen:2014wha}.
However, our Lagrangian arises as the $\mathcal O(1)$ part in the $c\rightarrow\infty$ limit of the relativistic Lagrangian
{\it after} divergences of $\mathcal O(c^2)$ canceled between the spin connection and the non-minimal coupling term.
More precisely, the non-relativistic Dirac operator $\Dnr$ in (\ref{eqn:massive-NR-Dirac}) was derived from
$\Dnr_{-1/4}$ in (\ref{eqn:Dnr-def})  following this procedure.
As such, two additional criteria must be met to obtain a Milne covariant operator and an invariant Lagrangian.
The first is that the divergences need to cancel for generic $n$, without restricting it to satisfy e.g.\ $n\wedge dn=0$.
The reason is that this condition is preserved by the transformation (\ref{eqn:Milne-shift-n}) only to leading order.
So if the divergences canceled for $n\wedge dn=0$ only, performing a shift of $n$ as in (\ref{eqn:Milne-shift-n}) would produce
$\mathcal O(1)$ pieces from the $O(c^2)$ divergences.
Since the full relativistic Lagrangian is invariant, these $\mathcal O(1)$ pieces would have to be canceled by what we defined as non-relativistic
Lagrangian and $L_\mathrm{nr}$ could not be invariant.
We carefully avoided using such restrictions in sec.~\ref{sec:NR-limit-general}, where we explicitly included acausal spacetimes with $n\wedge dn\neq 0$.
The second condition is that the identities which were used to show that the divergent parts cancel
must be preserved by the transformation (\ref{eqn:Milne-shift-n}), again also at subleading order in $c$.
This condition is independent from the first one, and we will see an example of how the non-relativistic Lagrangian
fails to be invariant if it is not met in sec.~\ref{sec:causal-spacetimes}.
The mechanism is the same as before: if the identities are not preserved exactly, the $\mathcal O(c^2)$ divergent terms
produce $\mathcal O(1)$ pieces when shifting $n$, and these have to be compensated by the transformation of $L_\mathrm{nr}$.
When these two criteria are met, our procedure gives an invariant Lagrangian.
For our construction, meeting the second condition just amounts to $ic\slashed{n}P_\pm=\pm P_\pm$
being preserved under (\ref{eqn:Milne-shift-n}), which is certainly the case.
Thus, we indeed expect $L_\mathrm{nr}$ to be Milne invariant.
However, a complete discussion of the symmetries certainly has to include the transformations of the individual building
blocks of the Lagrangian, and to derive these we go through the above arguments more explicitly now.

We start with the non-relativistic spinor $\Psi=\mathcal P\psi$.
Since $\psi$ does not transform under Milne boosts, the only transformation is that resulting from $\mathcal P$.
We find
\begin{align}\label{eqn:spinor-Milne}
 \Psi&\rightarrow \Psi+\frac{i}{2}\slashed{\zeta}P_+\Psi~,
 &\overline{\Psi}&\rightarrow\overline{\Psi}+\frac{i}{2}\overline{\Psi}P_+\slashed{\zeta}~.
\end{align}
To derive the transformation of $\Dnr$ in (\ref{eqn:massive-NR-Dirac}),
which defines the operator in terms of non-relativistic objects,
we start from $\Dnr_\alpha$ in (\ref{eqn:Dnr-def}). Namely,
\begin{align}\label{eqn:Dnr-def-copied}
 \Dnr_\alpha&=\Pt\left(\slashed{D}-mc -\frac{i\alpha}{2mc}\slashed{F}\right)\Pt\Big\vert_{c\rightarrow\infty}~.
\end{align}
Once again, the only objects transforming non-trivially under Milne boosts are the projectors $\tilde P$.
We find
\begin{align}\label{eqn:NR-Dirac-Milne}
 \Dnr_\alpha&\rightarrow \Big(\mathds{1}-\frac{i}{2}\slashed{\zeta}P_-\Big)\Dnr_\alpha\Big(\mathds{1}-\frac{i}{2}P_-\slashed{\zeta}\Big)~,
\end{align}
and $\Dnr_\alpha$ thus transforms covariantly for any $\alpha$.
To infer whether $\Dnr$ in (\ref{eqn:massive-NR-Dirac}) transforms covariantly,
we note that between $\Dnr_\alpha$ and $\Dnr$ we have canceled the divergences in (\ref{eqn:divergence-Omega-F}),
along with the mass term and the leading part of $C$.
We then need to check whether or not these divergent pieces contribute a non-trivial Milne variation at $\mathcal O(1)$.
We define a shorthand for the divergent pieces with general $\alpha$ as
\begin{subequations}\label{eqn:divergent-pieces-Milne}
\begin{align}
 \Dnr_{\alpha,\mathrm{div}}&:=\Pt \left(\slashed{D}-mc -\frac{i\alpha}{2mc}\slashed{F}\right)\Pt\,\Big\vert_{\mathcal O(c^2)}\\
 &\hphantom{:}=
 c^2P_+\left[
 \frac{1}{8}\partial_{[\mu}n_{\nu]}\left(c\slashed{n}\gamma^{\mu\nu}-3\gamma^{\mu\nu}c\slashed{n}+8i\alpha\gamma^{\mu\nu}\right)
 +imc\slashed{n} -m
 \right]P_+~.
\end{align}
\end{subequations}
The first term in square brackets combines the divergent pieces coming from the spin connection and the non-minimal coupling,
as given in eqn.\ (\ref{eqn:divergence-Omega-F}).
The second term is the divergent piece coming from the relativistic background gauge field $C$,
and the last one is the mass term.
We start with the terms containing $m$. After the Milne transformation they become
\begin{align}
 P^\prime_+ \left(imc\slashed{n}^\prime -m\right)P^\prime_+~.
\end{align}
The projectors shift in the same way as $\slashed{n}$ does,
so the identity $ic\slashed{n}P_\pm=\pm P_\pm$
which made them cancel in the first place also makes their Milne transformed versions cancel.
They thus do not contribute a Milne variation to the $\mathcal O(1)$ part.

For the spin connection part and the non-minimal coupling term we use exactly the same argument, and absorb $\slashed{n}^\prime$
into $P_+^\prime$. This yields
\begin{subequations}
\begin{align}\label{eqn:delta-Dnr-O1-0}
 \Dnr_{\alpha,\mathrm{div}}^\prime&=
 ic^2\left(\frac{1}{4}+\alpha\right)P_+^\prime\partial_{[\mu} n^\prime_{\nu]}\gamma^{\mu\nu}P_+^\prime
 \\
 &=\Dnr_{\alpha,\mathrm{div}}-i\left(\frac{1}{4}+\alpha\right)P_+\partial_{[\mu} \xi_{\nu]}\gamma^{\mu\nu}P_+
 +\dots~,
 \label{eqn:delta-Dnr-O1}
\end{align}
\end{subequations}
where the dots in the second line denote terms without derivatives of $\xi$.
From eqn.\ (\ref{eqn:delta-Dnr-O1-0}) we find that for $\alpha=-\frac{1}{4}$ the variations cancel exactly,
in addition to $\Dnr_{\alpha,\mathrm{div}}$ itself vanishing.
We also see now what the implications of that choice of $\alpha$ are when $n\wedge dn=0$.
In that case, $\Dnr_{\alpha,\mathrm{div}}=0$ for any choice of $\alpha$, as can be seen by using (\ref{eqn:zero-ndn-1}).
However, the Milne variation in (\ref{eqn:delta-Dnr-O1}) does {\it not} vanish:
recall that the dots denote terms without derivatives of $\xi$ and can thus not cancel the first term.
Since the entire $\Dnr_\alpha$ still is Milne covariant, this means that
$\Dnr$ transforms with the opposite non-covariant $d\xi$-term.
That is, for $\alpha\neq-\frac{1}{4}$ the non-relativistic Dirac operator $\Dnr$ defined in (\ref{eqn:massive-NR-Dirac}) transforms
non-covariantly even for $n\wedge dn=0$, e.g.\ on flat space with $n=dt$.

Coming back to $\alpha=-\frac{1}{4}$, eqn.\ (\ref{eqn:delta-Dnr-O1-0}) shows that the divergent pieces which we canceled
on the way from $\Dnr_\alpha$ to $\Dnr$ do not contribute a Milne variation.
With the analogous statement for the mass-like terms as shown above,
this completes the argument that $\Dnr$ indeed transforms as
\begin{align}\label{eqn:NR-Dirac-Milne-final}
 \Dnr&\rightarrow \Big(\mathds{1}-\frac{i}{2}\slashed{\zeta}P_-\Big)\Dnr\Big(\mathds{1}-\frac{i}{2}P_-\slashed{\zeta}\Big)~.
\end{align}
We see that the non-relativistic Dirac/L\'evy-Leblond operator transforms covariantly, and together with eqn.\ (\ref{eqn:spinor-Milne})
that our non-relativistic Lagrangian (\ref{eqn:NR-Lagrangian}) is thus Milne invariant.
As a final remark on Milne invariance, note that we needed $a$ in the non-relativistic Lagrangian and its Milne transformation.
This allowed us to separate the spin connection into leading pieces having only $n$ in eqn.\ (\ref{eqn:spin-connection-expansion}),
as opposed to the full $e^{\underline{0}}$, which is crucial for the cancellation of the Milne variations.

\subsection{Local Galilean transformations}\label{sec:symmetries-Galilean}
We now come to local Galilean invariance.
We will need the induced transformations for the spinor and vielbein, i.e.\ for spinor and vector representations.
The latter are straightforward, while the spinor transformations are a bit more subtle.
In the relativistic theory we want to look at a combination of local Lorentz transformations and
Milne boosts which leave $\slashed{n}$ and thus the projectors $P_\pm$ invariant.
This particular combination will turn out to induce local Galilean transformations on the non-relativistic objects.
Under a local Lorentz transformation we have $\slashed{n}\rightarrow\Lambda_s \slashed{n}\Lambda_s^{-1}$,
and to keep $\slashed{n}$ invariant, we will compensate with a Milne boost.%
\footnote{%
Milne and local Lorentz transformations leave the norm of $n_a$ invariant and, as we shall see shortly,
transform it by subleading terms only, so they can indeed compensate each other.
}
We start with a Lorentz transformation parametrized by $\tau$ as
$\Lambda_s=\operatorname{exp}(\tau_{ab}\Sigma^{ab})$,
where $\Sigma^{ab}$ is a basis of the Lie algebra $\mathfrak{so}(1,d\,{-}\,1)$. We have
\begin{align}
 \slashed{n}&\rightarrow e^{\tau_{ab}\Sigma^{ab}}\slashed{n}e^{-\tau_{ab}\Sigma^{ab}}
 =\slashed{n}+[\tau_{ab}\Sigma^{ab},\slashed{n}]+\dots~,
 &\Sigma^{ab}&=\frac{1}{2}\gamma^{ab}~.
\end{align}
We start with the infinitesimal transformation, the exponentiated version then works analogously.
The antisymmetric parameter matrix $\tau$ can be decomposed into
\begin{align}\label{eqn:Lorentz-trafo-split}
 \tilde\tau_{ab}&=\Pg{a}c\Pg{b}d\tau_{cd}~,&
 \tau_a&=2\tau_{ab}v^b~,
\end{align}
such that $\tau_{ab}\Sigma^{ab}=\tilde\tau_{ab}\Sigma^{ab}+\tau_a n_b\Sigma^{ab}$.
Note that $\tau_a v^a=0$.
Since $\tilde\tau_{ab}\Sigma^{ab}$ commutes with $\slashed{n}$, we have
\begin{align}\label{eqn:localLorentz-nslash}
 \delta_\mathrm{L}\slashed{n}&=[\tau_{ab}\Sigma^{ab},\slashed{n}]=
 [\tau_{a}\Sigma^{ab}n_b,\slashed{n}]
 =\frac{1}{2}[\tau_a\tilde\gamma^a\slashed{n},\slashed{n}]
 = -\frac{1}{c^2}\tau_a\tilde\gamma^a~.
\end{align}
To compensate for the change in $\slashed{n}$ due to the local Lorentz transformation $\delta_L$, we combine it
with a Milne boost under which $\delta_\mathrm{M}n_\mu=c^{-2}\tau_\mu$, where $\tau_\mu=2e_\mu^a\tau_{ab} v^b$ is spatial.
This works out just as well for the finite transformations.
For the Lorentz transformation we simply exponentiate $\tau_{ab}\Sigma^{ab}$.
For the Milne boost we use the notation of (\ref{eqn:Milne-shift-n}) and
set $\xi_\mu=\lambda n_\mu + \tau_\mu$ with $\lambda=c^2-c\sqrt{c^2+\tau^\mu\tau_\mu}$.
We note that only a subset of the quantities transforms under Milne boosts.
In particular, the vielbein is invariant and is thus only affected by the local Lorentz transformation.
But it is the particular combination of Milne and Lorentz transformations that leaves $\mathcal{P}$ and $\Pt$ invariant.

\subsubsection{Vector representation and spin connection}
We start by recapitulating how a local Lorentz transformation contracts to a local Galilean transformation for the vector representation.
We take a Lorentz transformation parametrized by an orthogonal matrix $\Lambda$
satisfying $\Lambda^T\eta\Lambda=\eta$, or with explicit indices
\begin{align}
 \Lambda^{b}_{\hphantom{b}a}\eta_{bc}\Lambda^{c}_{\hphantom{c}d}&=\eta_{ad}~.
\end{align}
Projecting this on the spatial components shows $n_a\Lambda^a_{\hphantom{a}c}\Pg{b}c\big\vert_{c\rightarrow\infty}=0$,
and we find
\begin{align}
 \Lambda^a_{\hphantom{a}b}\big\vert_{c\rightarrow\infty}&=\tilde\Lambda^a_{\hphantom{a}b}+\Lambda^a n_b~,
&\tilde \Lambda^a_{\hphantom{a}b}&=P^a_{\hphantom{a}c}\Lambda^c_{\hphantom{c}d}P^d_{\hphantom{d}b}~,
&\Lambda^a&=\Lambda^a_{\hphantom{a}b}v^b~.
\end{align}
To make the structure more apparent, we can fix $e_\mu^a$ such that $n_a=\delta_{a\underline{0}}$,
and then write $\Lambda^a_{\hphantom{a}b}\big\vert_{c\rightarrow\infty}$ in matrix form as
\begin{align}
 \Lambda^a_{\hphantom{a}b}\big\vert_{c\rightarrow\infty}&=\begin{pmatrix}
                                                           1 & 0\\
                                                           \Lambda^i  & \tilde \Lambda^i_{\hphantom{i}j}
                                                          \end{pmatrix}~.
\end{align}
This is the fundamental linear representation of the Galilean group.
We can sharpen the fall-off behavior of $\Lambda^{\underline{0}}_{\hphantom{\underline{0}}b}$ as follows.
From $g=e^{\underline{0}}_\mu e^{\underline{0}}_\nu + e_{\mu i}e^i_\nu=e^{\prime\underline{0}}_\mu e^{\prime \underline{0}}_\nu + e^\prime_{\mu i}e^{\prime i}_\nu$,
we see that boosts only transform the $\mathcal O(c^{-2})$ part of $e^{\underline{0}}$, i.e.\ $a$ in the notation of eqn.\ (\ref{eqn:e0-decomposition}).
Note, however, that under the particular linear combination of local Lorentz and Milne transformations we are looking at here
(which will give local Galilean transformations), $n_\mu$ and $e^{\underline{0}}$ transform in the same way, and thus $a$ is invariant.
This is different from \cite{Geracie:2015dea} --
without gauge fixing $e_{\underline{0}}^\mu=v^\mu$, we have the Milne and local Galilean transformations disentangled.

We now come to the spin connection $\hat\omega$ defined in eqn.\ (\ref{eqn:spin-connection-hat}).
To make its structure as a Lie-algebra-valued one-form explicit, we look at
\begin{align}\label{eqn:omega-dot-converted}
 \hat\omega_{\mu\hphantom{a}b}^{\hphantom{\mu}a}&=\eta^{ac}e^\nu_c\hat\omega_{\mu\nu\rho}e^\rho_b\big\vert_{c\rightarrow\infty}~.
\end{align}
The only explicit $c$-dependence on the right hand side of eqn.\ (\ref{eqn:omega-dot-converted}) is in $\eta$.
Due to $n_a\eta^{ac}=\mathcal O(c^{-2})$, only $P^a_{\hphantom{a}c} \hat\omega_{\mu\hphantom{c}b}^{\hphantom{\mu}c}$ has non-vanishing components at $c\rightarrow\infty$.
More explicitly,
\begin{align}\label{eqn:spin-connection-limit}
 \hat\omega_{\mu\hphantom{a}b}^{\hphantom{\mu}a}
 &=\hat\omega_{\mu\nu\rho}\,\tilde e^{\nu a}\tilde e^\rho_b
   +\hat\omega_{\mu\nu\rho}\,\tilde e^{\nu a}v^\rho n_b
 =:\tilde\omega_{\mu\hphantom{a}b}^{\hphantom{\mu}a}+\varpi_{\mu}^{\hphantom{\mu}a} n_b~.
\end{align}
This defines a spin connection for spatial rotations $\tilde\omega$,
and $\varpi$ is the boost connection.
We note that $n_a\varpi_\mu^{\hphantom{\mu}a}=0$.
To make the structure more apparent, we can write eqn.\ (\ref{eqn:spin-connection-limit}) for $n_a=\delta_{a\underline{0}}$ as
\begin{align}
 \hat\omega_{\mu\hphantom{a}b}^{\hphantom{\mu}a}&=
 \begin{pmatrix}
   0 & 0 \\ \varpi_{\mu}^{\hphantom{\mu}i} & \tilde\omega_{\mu\hphantom{i}j}^{\hphantom{\mu}i}
 \end{pmatrix}~.
\end{align}
This structure is exactly what we expect for a one-form with values in the Lie algebra of the Galilei group.
It immediately implies $D_\mu n_a=\partial_\mu n_a+\hat\omega_{\mu\hphantom{b}a}^{\hphantom{\mu}b}n_b=0$.
We also have $D_\mu h^{ab}=0$, which makes our connection satisfy the analog of metric compatibility spelled out in \cite{Geracie:2015dea}.
The non-relativistic Christoffel symbols $\dot\Gamma_{\mu\nu}^\rho$ can then be defined from $D_\mu e_\nu^a=0$.

\subsubsection{The spinor}
To justify calling $\Psi$ defined in (\ref{eqn:NR-spinor-def}) a non-relativistic spinor,
we have to show that it indeed carries a spin representation of the Galilean group.
We restrict the discussion to infinitesimal transformations, the exponentiated ones again follow a similar logic.
As explained above, we take a local Lorentz transformation accompanied by a Milne boost such that the projectors $P_\pm$ are invariant.
The relativistic spinor $\psi$ is Milne invariant, so
the transformation $\psi\rightarrow \psi+\delta \psi$ is solely due to the Lorentz part
\begin{align}
 \delta\psi&=\tau_{ab}\Sigma^{ab}~,&
 \Sigma^{ab}&=\frac{1}{2}\gamma^{ab}~.
\end{align}
The transformation of $\Psi$ can be derived using (\ref{eqn:NR-spinor-def}).
Separating the time and spatial components using (\ref{eqn:Lorentz-trafo-split}),
we find
\begin{align}\label{eqn:spinor-Galilean-transf0}
 \delta\Psi&=\tilde\tau^{}_{ab}\tilde\Sigma^{ab}\Psi
 -i \tau^{}_{a}\tilde\gamma^{a}\left(c^{-\frac{1}{2}}P_+ -c^{-\frac{3}{2}}P_-\right)\psi\big\vert_{c\rightarrow\infty}~,
 &\tilde\Sigma^{ab}&=\Pg{c}a\Pg{d}b \Sigma^{cd}~.
\end{align}
To get the second term, we have used that $P_\pm\Sigma^{ab}n_b=\Sigma^{ab}n_bP_\mp$,
along with $i\slashed{n}P_\pm=\pm c^{-1}P_\pm$.
Note how this mixes the orders in $c$, and exchanges the projectors for the boosts.\footnote{%
Note also how it was important to keep the projectors invariant. Otherwise we would get
$\Psi\rightarrow\Lambda_s\mathcal P\Lambda_s^{-1}\Lambda_s\psi=\Lambda_s\Psi$.
This would give {\it only} the spatial rotations at $c\rightarrow\infty$, as $\tau_a\tilde\gamma^a$ looses
one power in $c$ without exchanging projectors.
}
Evaluating eqn.\ (\ref{eqn:spinor-Galilean-transf0}) in the $c\rightarrow\infty$ limit leaves only the first term in the parentheses, and we find
\begin{align}\label{eqn:spinor-Galilean-transf}
 \delta\Psi&=\left[\tilde\tau^{}_{ab}\tilde\Sigma^{ab} - \tau^{}_{a}\Sigma^a\right]\Psi~,&
 \Sigma^a&=\frac{i}{2}\tilde\gamma^aP_+~.
\end{align}
The $\tilde\Sigma^{ab}$ generate the Lie algebra of SO($d\,{-}\,1$).
Furthermore, due to $P_\pm\tilde\gamma^{a}=\tilde\gamma^a P_\mp$, the $\Sigma^a$'s commute with each other.
The remaining commutator evaluates to
\begin{align}
 \left[\tilde\Sigma^{ab},\Sigma^c\right]&=h^{bc}\Sigma^a-h^{ac}\Sigma^b~.
\end{align}
That is precisely the Galilean Lie algebra, and we have thus obtained the Galiliean spin representation, induced
by the local Lorentz transformations of $\psi$ on our spinor $\Psi$ as defined in (\ref{eqn:NR-spinor-def}).
This indeed justifies calling it a non-relativistic spinor.

Note that for the construction of that spin representation we did not go through constructing
a Clifford algebra for a degenerate space (the tangent space with the inverse spatial metric $h^{\mu\nu}$).
This is different from the approach of \cite{Brooke:1978tr,Kunzle:1984dt}.
What we called $\dot\gamma^\mu$ or $v^\mu P_+$ does not square to zero but to itself.
The crucial point, however, is that $\Sigma^a$ still squares to zero and satisfies the commutator relations
that the Galilean boost generator satisfies in \cite{Brooke:1978tr,Kunzle:1984dt}.

\subsubsection{Dirac operator}
The remaining task is to show that the non-relativistic Dirac operator $\Dnr$
transforms covariantly under local Galilean transformations.
To that end, $\Dnr_\alpha$ in (\ref{eqn:Dnr-def}) again is a good starting point.
We introduce a shorthand $X$ and write
\begin{align}
 \Dnr_\alpha&=\Pt X \Pt\Big\vert_{c\rightarrow\infty}~,&
 X&=\slashed{D}-mc -\frac{i\alpha}{2mc}\slashed{F}~.
\end{align}
$X$ is invariant under Milne boosts, and
since the relativistic covariant derivative transforms covariantly
under local Lorentz transformations,
$\delta \slashed{D}=[\tau_{ab}\Sigma^{ab},\slashed{D}]$,
we have
$\delta X=[\tau_{ab}\Sigma^{ab},X]$.
We thus find
\begin{align}\label{eqn:Dnr-localGal}
 \delta\Dnr_\alpha&=\Pt [\tau_{ab}\Sigma^{ab},X] \Pt\Big\vert_{c\rightarrow\infty}
 =[\tilde\tau_{ab}\tilde\Sigma^{ab},\Dnr_\alpha]
   + \tau_a\overline{\Sigma}^a \Dnr_\alpha + \Dnr_\alpha \tau_a\Sigma^a~,
\end{align}
where $\overline{\Sigma}^a=\gamma^{\underline{0}}(\Sigma^a)^\dagger\gamma_{\underline{0}}$.
The transformation (\ref{eqn:Dnr-localGal}) is precisely how a non-relativistic Dirac operator should change under local Galilean transformations to get an invariant Lagrangian,
noting that $\delta\overline{\Psi}=-\overline{\Psi}(\tau_a\overline{\Sigma}^a+\tilde\tau_{ab}\tilde\Sigma^{ab})$.

To show that $\Dnr$ transforms correctly
we once again have to check that the transformations of the divergent pieces canceled on the way from $\Dnr_\alpha$ to $\Dnr$
do not pollute the $\mathcal O(1)$ part.
This boils down to almost the same argument as given in sec. \ref{sec:milneinvofL} for the Milne boosts.
This time, however, $\slashed{n}$ and $P_+$ are both invariant.
The analog of (\ref{eqn:delta-Dnr-O1-0}), i.e.\ the transformation of the divergent parts in $\Dnr_\alpha$,
but now under the particular combination of Milne boost
and local Lorentz transformation, reads
\begin{align}\label{eqn:delta-Dnr-div-Gal}
 \Dnr_{\alpha,\mathrm{div}}^\prime&=
 ic^2\left(\frac{1}{4}+\alpha\right)P_+\partial_{[\mu} n^\prime_{\nu]}\gamma^{\prime\mu\nu}P_+~.
\end{align}
Note that $P_+$ and $\slashed{n}$ do not transform, but $n_\mu$ and the inverse vielbeine in $\gamma^{\mu\nu}$ do.
We see that the variations cancel just in the same way as they did for Milne boosts when $\alpha=-\frac{1}{4}$ --
the point is that $P_\pm$ and $\slashed{n}$ still transform the same way.
For $\alpha\neq-\frac{1}{4}$, $\partial_{[\mu}n^\prime_{\nu]}$ again produces a $d\xi$ term at $\mathcal O(1)$, even if $n\wedge dn=0$.
So we once again need $\alpha=-\frac{1}{4}$, and then find
\begin{align}\label{eqn:Dnr-localGal-final}
 \delta\Dnr&=
   [\tilde\tau_{ab}\tilde\Sigma^{ab},\Dnr]
   + \tau_a\overline{\Sigma}^a \Dnr + \Dnr \tau_a\Sigma^a~,
\end{align}
as needed to have the non-relativistic Lagrangian (\ref{eqn:NR-Lagrangian}) invariant.

We note that
$(\dot\gamma^\mu+\tilde\gamma^\mu)\mathcal D_\mu$ alone transforms covariantly under spatial rotations.
This is because the additional pieces in $\Dnr$ transform covariantly by themselves, and spatial rotations
do not alter the cancellation of the $n\wedge dn$ divergent pieces.
However, for the local Galilean boosts which arise as combination of local Lorentz and Milne transformations,
we do require the non-minimal coupling term to get covariant transformation of the $\mathcal O(1)$ part alone.

In summary, we have shown that each of the pieces in the non-relativistic Lagrangian (\ref{eqn:NR-Lagrangian}) transforms covariantly under
local Galilean transformations, and we thus have an invariant Lagrangian.
The local Galilean transformations arise from the relativistic theory as combination of Milne boosts and local Lorentz transformations such
that $\slashed{n}$ and thus the projectors $P_\pm$ are invariant.
This combination is different from Milne boosts alone:
the latter act on the ``metric data'' and leave the vielbein invariant, while the former act on the vielbein and ``frame data''.
We note that these statements depart from those in \cite{Geracie:2015dea}, where Milne transformations were identified with Galilean boosts.
Having Milne and Galilean transformations as separate invariances seems to make sense from a general perspective:
Milne boosts reflect the gauge redundancy in splitting the cotangent space into time and spatial directions, via $v$,
while the local Galilean symmetries reflect the gauge redundancy in choosing a section in the frame bundle,
which is a different geometric object. 
Technically, the reason we can see the two invariances separately is that we did not gauge fix $e_{\underline{0}}^\mu=v^\mu$.
Upon imposing that choice, our constructions for the frame data agree with~\cite{Geracie:2015dea}.

\section{Causal spacetimes}\label{sec:causal-spacetimes}

We now specialize to causal spacetimes and derive a Lagrangian for the dynamical fields alone from the general non-relativistic Lagrangian (\ref{eqn:NR-Lagrangian}).
By causal spacetimes we mean Newton-Cartan backgrounds with $n\wedge dn=0$, ensuring that a foliation in terms of constant-time hypersurfaces exists.
As seen in eqn.\ (\ref{eqn:massive-NR-Dirac}), the equation of motion for $\Psi_-:= P_-\Psi$ does not have a time derivative and is a constraint equation.
One can then integrate $\Psi_-$ out and write a Lagrangian purely in terms of $\Psi_+ := P_+\Psi$.
This obscures the structure and part of the symmetries but allows for easier comparison to some of the existing literature.

To facilitate the computations, we split the derivative part of $\Dnr$ as
\begin{align}
 \left(\dot\gamma^\mu+\tilde\gamma^\mu\right)\mathcal D_\mu&=:\slashed{D}_s+\slashed{D}_t~,
\end{align}
such that $[\slashed{D}_t,\slashed{n}]=0$ and $\lbrace\slashed{D}_s,\slashed{n}\rbrace=0$, and thus $[\slashed{D}_t,P_\pm]=0$ and $\slashed{D}_s P_\pm = P_\mp\slashed{D}_s$.
The explicit expressions are
\begin{align}
 \slashed{D}_\mathrm{s}&=\tilde\gamma^\mu\left[\partial_\mu+\frac{1}{4}\hat\omega_{\mu\nu\rho}\tilde\gamma^{\nu\rho}-im A_\mu\right]~,&
 \slashed{D}_\mathrm{t}&=\dot\gamma^\mu\left[\partial_\mu +\frac{1}{4}(\hat\omega_{\mu\nu\rho}+2\hat\omega_{\nu\rho\mu})\tilde\gamma^\nu\tilde\gamma^\rho
   -im A_\mu\right]~.
\end{align}
These are defined out of convenience for the split into ``chiral'' components,
and we leave a check of their transformation properties for the future.
So far with no extra assumptions, we then find for the non-relativistic Lagrangian (\ref{eqn:NR-Lagrangian})
\begin{align}\label{eqn:NR-Lagrangian-split}
\begin{split}
 L_\mathrm{nr}=\,&
 i\overline{\Psi}_+\Big[\slashed{D}_\mathrm{t}+\frac{i}{8}\slashed{F}^{(A)}\Big]\Psi_+
 +i\overline{\Psi}_-\Big[\slashed{D}_\mathrm{s}+\frac{1}{4}T^\mathrm{c}_{\mu}\tilde\gamma^\mu\Big]\Psi_+
 \\&
 +i\overline{\Psi}_+\Big[\slashed{D}_\mathrm{s}+\frac{3}{4}T^\mathrm{c}_{\mu}\tilde\gamma^\mu\Big]\Psi_-
 -i\overline{\Psi}_-\Big[2m+\frac{i}{4}\slashed{T}^\mathrm{c}\Big]\Psi_-
 +\cc~,
\end{split}
\end{align}
where
$\slashed{F}^{(A)}=F^{(A)}_{\mu\nu}\tilde\gamma^{\mu\nu}$,
analogously
$\slashed{T}^\mathrm{c}=T^\mathrm{c}_{\mu\nu}\tilde\gamma^{\mu\nu}$,
and
$T_\mu^\mathrm{c}=T^\mathrm{c}_{\mu\nu}v^\nu$.
We now exploit the simplifications that are specific to causal spacetimes.
From eqn.\ (\ref{eqn:zero-ndn-1}), we see that the torsion coupling $\overline{\Psi}_-\slashed{T}^\mathrm{c}\Psi_-$
in (\ref{eqn:NR-Lagrangian-split}) drops out in causal spacetimes.
We can then solve the $\Psi_-$ equation of motion for $\Psi_-$, which yields
\begin{align}
 \Psi_-&=\frac{1}{2m}\left[\slashed{D}_\mathrm{s}+\frac{1}{4}T_\mu^\mathrm{c}\tilde\gamma^\mu\right]\Psi_+~.
\end{align}
Substituting this back into the Lagrangian (\ref{eqn:NR-Lagrangian-split}), we find an expression purely in terms of $\Psi_+$.
Namely,
\begin{align}\label{eqn:NR-Lagrangian-psi-plus}
 L_\mathrm{nr}=\,&
 i\overline{\Psi}_+\Big[\slashed{D}_\mathrm{t}+\frac{i}{8}\slashed{F}^{(A)}\Big]\Psi_+
 +\frac{i}{2m}\overline{\Psi}_+\Big[\slashed{D}_\mathrm{s}+\frac{3}{4}T_\mu^\mathrm{c}\tilde\gamma^\mu\Big]
 \Big[\slashed{D}_\mathrm{s}+\frac{1}{4}T_\mu^\mathrm{c}\tilde\gamma^\mu\Big]\Psi_+
 +\cc~.
\end{align}
We note that the ``chirality'' of $\Psi_+$ is not preserved by local Galilean boosts or Milne boosts.
This can be seen explicitly from eqns.\ (\ref{eqn:spinor-Milne}) and (\ref{eqn:spinor-Galilean-transf}), which show that either of the boosts produces
a negative chirality piece from a positive chirality spinor.
Thus, the split into chiralities depends on the chosen frame, which
makes the full non-relativistic Lagrangian (\ref{eqn:NR-Lagrangian}) more appealing.
What remains a symmetry of (\ref{eqn:NR-Lagrangian-psi-plus}) is a combination of local Galilean transformations and Milne boosts,
where $\zeta_\mu$, parametrizing the Milne boost, is related to $\tau_a$ parametrizing a local Galilean boost,
such that $\psi$ is invariant.\footnote{%
In the language of sec.~\ref{sec:milneinvofL}, the reason for not having all symmetries realized is that the second criterion
is not met:
disposing of $\Psi_-$, to have a Lagrangian in terms of $\Psi_+$ only,
leaves $\Psi_+$ as just a spinor with half as many components, as opposed to being defined from
$\Psi$ with an explicit projector $P_+$ as before.
With the projector implicit, the identity $ic\slashed{n}\Psi_+=\Psi_+$, which we need to show that the divergences cancel,
is not preserved by~(\ref{eqn:Milne-shift-n}).
\label{foot:chirality}
}

\subsection{Implications for flat space}\label{sec:flat-space}
As an illustrative example, we spell out the non-relativistic Lagrangian and Dirac operator explicitly for a flat Newton-Cartan spacetime,
by which we mean $n=dt$ and $h^{-1}=\delta^{ij}\partial_i\otimes\partial_j$.
To get this structure from a non-relativistic limit we choose $g=-c^2dt^2+d\vec{x}^2$ and $n=dt$.
We also fix the vielbein to $e_\mu^a=\delta_\mu^a$, which completely fixes the local Lorentz symmetry in the relativistic
theory and correspondingly the local Galilean symmetry in the non-relativistic limit.
The global symmetries are those combinations of diffeomorphisms and Galilean transformations which leave the Newton-Cartan structure
and the chosen frame invariant.
These choices also fix $a=0$ from (\ref{eqn:e0-decomposition}).
The non-relativistic Lagrangian (\ref{eqn:NR-Lagrangian}) becomes
\begin{align}\label{eqn:flat-space-Lagrangian}
 L_\mathrm{nr}&=i\overline{\Psi}\left[\left(\dot\gamma^\mu+\tilde\gamma^\mu\right)\left(\partial_\mu-i m A_\mu\right)-2m P_-
 +\frac{i}{8}F_{\mu\nu}^{(A)}\tilde\gamma^{\mu\nu}P_+\right]\Psi+\cc ~.
\end{align}
We see that, not quite surprisingly, all the extra torsion couplings have dropped out.
The non-minimal coupling to the background gauge field, however, is still present.
Solving for $\Psi_-$ once again yields eqn.\ (\ref{eqn:NR-Lagrangian-psi-plus}), which for the flat case reads
\begin{align}
 L_\mathrm{nr}=\,&
 \overline{\Psi}_+\Big[v^\mu\left(\partial_\mu-im A_\mu\right)-\frac{1}{8}\slashed{F}^{(A)}\Big]\Psi_+
 +\frac{i}{2m}\overline{\Psi}_+\Big[\tilde\gamma^\mu\left(\partial_\mu-im A_\mu\right)\Big]^2\Psi_+
 +\cc~.
\end{align}
Note that the coupling to $\slashed{F}^{(A)}$ in the first term is the result of the
non-minimal coupling term in (\ref{eqn:non-minimally-coupled-Lagrangian}) with $\alpha=-\frac{1}{4}$.
Evaluating the squared spatial Dirac operator in the second term yields, as usual, a Laplace operator (the spatial one in this case)
and a coupling to the gauge field strength
\begin{align}\label{eqn:spatial-dirac-squared}
 \left[\tilde\gamma^\mu\left(\partial_\mu-im A_\mu\right)\right]^2
 &=
 \bigtriangleup_\mathrm{s} - \frac{im}{2}\tilde\gamma^{\mu\nu}F_{\mu\nu}^{(A)}~,
 &
 \bigtriangleup_\mathrm{s}&=h^{\mu\nu}(\partial_\mu-imA_\mu)(\partial_\nu-imA_\nu)~.
\end{align}
The analog in curved backgrounds would have additional curvature and torsion couplings.\footnote{%
In fact, one would conveniently define the curvatures from this sort of relation. That is, write
$\Dnr=:(\dot\gamma^\mu+\tilde\gamma^\mu)\hat{\mathcal D}_\mu$ and then define
spatial and boost curvature and torsion along the lines of
 $[\hat {\mathcal D}_\mu,\hat {\mathcal D}_\nu]=:R_{\mu\nu}^{ab}\tilde\Sigma_{ab}+R_{\mu\nu}^a\Sigma_a+T_{\mu\nu}^a(\tilde\gamma_a+\dot\gamma_a)$.
}
In flat space the non-relativistic Lagrangian simply becomes
\begin{align}
 L_\mathrm{nr}&=\overline{\Psi}_+v^\mu\left(\partial_\mu-im A_\mu\right) \Psi_+ +\frac{i}{2m}\overline{\Psi}_+\bigtriangleup_\mathrm{s} \Psi_+
 + \frac{1}{8}\overline{\Psi}_+\tilde\gamma^{\mu\nu}F^{(A)}_{\mu\nu}\Psi_+ +\cc~.
\end{align}
The crucial feature of this Lagrangian is the magnetic coupling to $\tilde\gamma^{\mu\nu}F^{(A)}_{\mu\nu}$.
Had we started with just a free Dirac field, without the non-minimal coupling term in eqn.\ (\ref{eqn:non-minimally-coupled-Lagrangian}),
the coefficient of $\overline{\Psi}_+\tilde\gamma^{\mu\nu}F^{(A)}_{\mu\nu}\Psi_+$ would be $\frac{1}{4}$,
which is just the contribution from eqn.\ (\ref{eqn:spatial-dirac-squared}) corresponding to the familiar relativistic gyromagnetic ratio $g\,{=}\,2$.
The effect of the non-minimal coupling with $\alpha=-\frac{1}{4}$ in the non-relativistic theory is to reduce this coupling to
$\frac{1}{8}$, corresponding to $g\,{=}\,1$.
We thus explicitly see how demanding either consistency in generic backgrounds
or Milne invariance in any background has striking implications
even for flat space.
We note that the preferred value of $g$ for which we find the full symmetry realized is precisely $g=2s$, where $s=\frac{1}{2}$ is the spin.

\section{Three-dimensional spacetimes}\label{sec:3d}
In sec.~\ref{sec:NR-limit-general} we saw that the divergences in the spin connection part can not in general be canceled
by a modification of the gauge field (\ref{eqn:gauge-field-expansion}), and we thus needed the non-minimal coupling term with $\alpha=-\frac{1}{4}$.
This was because the divergences in $\slashed{C}$ and $\Omega$ have different Clifford algebra structures.
The antisymmetric product of $k$ $\gamma$-matrices is generically related to an antisymmetric product of $d\,{-}\,k$ ones
 in $d$ dimensions, by a form of Hodge duality.
This turns out to make a difference precisely for $d=3$.
In our conventions we have
\begin{align}\label{eqn:3d-gamma-duality}
 \gamma_{\mu\nu\rho}&=\pm c\epsilon_{\mu\nu\rho}\mathds{1}~,
\end{align}
where $\epsilon_{\mu\nu\rho}=\det(e)\, \varepsilon_{\mu\nu\rho}$ and $\varepsilon_{\mu\nu\rho}=\pm 1,0$.
Note that $\det(e)=\sqrt{\det\upgamma}$, where $\upgamma_{\mu\nu}=n_\mu n_\nu+h_{\mu\nu}$ was used to define the volume form in sec.~\ref{sec:NR-limit-general}.
Therefore, $\epsilon_{\mu\nu\rho}$ is $\mathcal O(1)$ at $c\rightarrow\infty$ and we have an
explicit factor of $c$ on the right hand side of eqn.\ (\ref{eqn:3d-gamma-duality}).
The sign is a matter of convention and we pick the positive one.
We then immediately get $\gamma^{\mu\nu}=c \epsilon^{\mu\nu\rho}\gamma_\rho$.
With this identity we can further evaluate the divergence in the spin connection part included in eqn.\ (\ref{eqn:divergence-Omega-F}),
which in $d\,{=}\,3$ becomes
\begin{align}\label{eqn:divergence-Omega-F-3d}
 \Pt\Omega\Pt\Big\vert_{\mathcal O(c^2)}&=
 \frac{ic^2}{4}(\star dn)_\mu \,P_+c\gamma^\mu P_+~,
 &
 (\star dn)_\mu&=\epsilon_\mu^{\hphantom{\mu}\nu\rho}\partial_{\nu} n_{\rho}~.
\end{align}
The indices of $\epsilon_\mu^{\hphantom{\mu}\nu\rho}$ in the definition of $\star dn$ are raised with the relativistic metric.
This divergence has the same Clifford algebra structure as the terms resulting from the $\slashed{C}$ term in the relativistic Lagrangian,
and modifying $C$ thus is an additional option now to cancel the divergences.
There are two candidate terms, and our ansatz for the gauge field is
\begin{align}\label{eqn:gauge-field-expansion-3d}
 C_\mu&=
 -m c^2 \left(n_\mu +f_1(\star dn)_\mu + f_2 \mathcal B n_\mu\right)
 + m A_\mu~,
 &
 \mathcal B&=\frac{1}{\det e}\varepsilon^{\alpha\beta\gamma}n_\alpha\partial_\beta n_\gamma~,
\end{align}
where $f_1$ and $f_2$ are functions of $n$ and its derivatives.
Note that both extra terms agree at leading order in $c$, due to $v^\mu(\star dn)_\mu=-c^2\epsilon^{\alpha\beta\gamma}n_\alpha\partial_\beta n_\gamma=\mathcal B$.
As far as canceling the divergent piece is concerned, they are thus equally useful.
However, they are different in the subleading orders.
The form of the non-relativistic Lagrangian will depend on the relative strength of $f_1$ and $f_2$,
but the difference merely corresponds to a redefinition of the non-relativistic gauge field.
This can be seen by noting that, with $f_+:=f_1+f_2$ and $\mathcal T_\mu:=\epsilon_{\mu\nu\rho}v^\rho h^{\nu\sigma}T^\mathrm{c}_{\sigma}$,
\begin{align}
 C_\mu&=
 -m c^2 n_\mu\left(1 + f_+ \mathcal B \right)
 + m \mathcal A_\mu~,
 &
 \mathcal A_\mu&=A_\mu+f_1\mathcal T_\mu~.
\end{align}
Note that $v^\mu\mathcal T_\mu=0$.
We will include the non-minimal coupling term with generic $\alpha$ and determine $f_+$ from the requirement
that $\Pt (\slashed{D}-mc -\frac{i\alpha}{2mc}\slashed{F})\Pt$ be finite.
To find the non-relativistic Dirac operator with these extra parameters, we need to calculate the individual pieces again.
For the spin connection part we find, using $\tilde\gamma^{\mu\nu}T_{\mu\nu}^c=2\mathcal B c\slashed{n}$,
\begin{align}
 \Pt \Omega\Pt=\,&
 \frac{1}{4}\mathcal B (c^2P_+ + P_-)
 +\left(\dot\gamma^\mu+\tilde\gamma^\mu\right)\frac{1}{4}\dot\omega_{\mu\nu\rho}\left(\tilde\gamma^{\nu\rho}+2\tilde\gamma^\nu\dot\gamma^\rho\right)
 +\frac{1}{2}\tilde\gamma^\mu T_\mu^\mathrm{c}~,
\end{align}
Here and in the next two equations, contributions which vanish when $c\rightarrow\infty$ have been dropped.
For the remaining parts of the derivative we find
\begin{align}
\begin{split}
  \Pt \left[\gamma^\mu\left(\partial_\mu-i C_\mu\right)\right]\Pt
  =\,&
  \left(\dot\gamma^\mu+\tilde\gamma^\mu\right)\Big(\partial_\mu+\frac{1}{2}(\partial_\mu a_b)e_\nu^b n_\rho\tilde \gamma^\nu\dot\gamma^\rho-i m A_\mu\Big)
  \\&
   +m (1+f_+\mathcal B) \left(c^2P_+-P_-\right)~.
\end{split}
\end{align}
Evaluating the mass term is straightforward, and the only non-trivial information we still need is the non-minimal coupling term,
which evaluates to
\begin{align}
\begin{split}
 \Pt\Big(-\frac{i\alpha}{2mc}\slashed{F}\Big)\Pt&=
 \alpha \mathcal B\left(1+f_+\mathcal B\right)\left(c^2P_+-P_-\right)
 +i\alpha\left(1+f_+\mathcal B\right)T_{\mu\nu}^c\left(\dot\gamma^\mu\tilde\gamma^\nu+\tilde\gamma^\mu\dot\gamma^\nu\right)
 \\&\hphantom{=}\,
 -\frac{i\alpha}{2}\slashed{F}^{(\mathcal A)}P_+
 +\alpha\tilde\gamma^\mu(P_+-P_-)\partial_\mu(f_+\mathcal B)~.
\end{split}
\end{align}
Note that $F^{(\mathcal A)}=d\mathcal A$ is the field strength of the modified gauge field $\mathcal A$.
Demanding the divergent pieces in $\Pt (\slashed{D}-mc -\frac{i\alpha}{2mc}\slashed{F})\Pt$ to cancel
can now be seen to imply
\begin{align}\label{eqn:f1-f2-constraint}
 f_+&=-\frac{\alpha+\frac{1}{4}}{m+\alpha \mathcal B}~.
\end{align}
With that relation we see that the divergences cancel and that there are no $\mathcal O(c)$ subleading divergences.
Using eqn. (\ref{eqn:f1-f2-constraint}) we can assemble the non-relativistic Dirac operator $\Dnr$,
which now is $\Dnr_{\alpha}$ with generic $\alpha$, expressed in terms of Newton-Cartan data.
This yields
\begin{align}\label{eqn:massive-NR-Dirac-3d}
\begin{split}
 \Dnr&=\left(\dot\gamma^\mu+\tilde\gamma^\mu\right) \mathcal D^{(\mathcal A)}_\mu- 2mP_-
  -\frac{i\alpha}{2}\slashed{F}^{(\mathcal A)}P_+
  +\frac{1}{2}\tilde\gamma^\mu T_\mu^\mathrm{c}
  +\frac{1}{2}\mathcal BP_-
  \\&\hphantom{=}\,
  +i\alpha (1+f_+\mathcal B) T_{\mu\nu}^c\left(\dot\gamma^\mu\tilde\gamma^\nu+\tilde\gamma^\mu\dot\gamma^\nu\right)
  +\alpha\tilde\gamma^\mu(P_+-P_-)\partial_\mu(f_+\mathcal B)~,
\end{split}
\end{align}
where the superscripts on $\mathcal D_\mu$ and $F$ indicate that we have replaced
$A_\mu$ by $\mathcal A_\mu$.
The derivative $\mathcal D_\mu$ was defined in eqn.\ (\ref{eqn:calD-mu-def}),
and the slashed quantities below eqn.\ (\ref{eqn:NR-Lagrangian-split}).
On causal spacetimes, where $\mathcal B=0$, we have $-mf_+=\alpha+\frac{1}{4}=:\alpha_+$.
The non-relativistic Lagrangian once again is
\begin{align}\label{eqn:NR-Lagrangian-3d}
 L_\mathrm{nr}=i\overline{\Psi}\Dnr\Psi+\cc~.
\end{align}
The arguments for invariance under local Galilean transformations and Milne boosts proceed along the lines given in sec.~\ref{sec:symmetries}.
The divergent pieces cancel again exactly, leaving the symmetry transformations induced from the relativistic ones as symmetries
of the non-relativistic Lagrangian.
We will see below, however, that the transformations are altered.
We have thus obtained the fully covariant version of the Lagrangian studied in \cite{2015PhRvD..91d5030G},
where e.g.\ a Milne frame was fixed
and local Galilean invariance was not obvious.
To complete the discussion, we note that $\alpha_+$ is related to the gyromagnetic ratio $g$ by $4\alpha_+=g-1$.
On spacetimes with $n\wedge dn=0$ we can integrate out the $\Psi_-$ part of the two-component spinor $\Psi$, as shown in the previous section,
to obtain a Lagrangian for the one-component field $\Psi_+$ only.
This is the language used in \cite{2015PhRvD..91d5030G}.

We now come to the transformation properties under Milne boosts and local Galilean transformations.
The only quantity for which the transformation changes is the non-relativistic gauge field $A$.
Due to the difference in the leading term in eqn.\ (\ref{eqn:gauge-field-expansion-3d}) as compared to eqn.\ (\ref{eqn:gauge-field-expansion}),
the shifts of $n$ by a subleading part according to eqn.\ (\ref{eqn:Milne-shift-n}) result in a different behavior.
To match the transformations to \cite{Jensen:2014wha}, we set $\mathcal B=0$ {\it after} the variations are performed.
We then find
\begin{align}\label{eqn:anomalous-Milne-A-3d}
 A_\mu&\rightarrow A_\mu -\xi_\mu - f_2 n_\mu\epsilon^{\nu\rho\sigma}\partial_\nu(n_\rho\xi_\sigma)-(f_1+2f_2)n_\mu \epsilon^{\nu\rho\sigma}n_\nu\partial_\rho\xi_\sigma~.
\end{align}
This reduces to $A\rightarrow A-\xi$ as given in eqn.\ (\ref{eqn:aA-Milne-trafo}) for $\alpha_+=f_1=f_2=0$, or $g=1$.
Upon identifying $g_1$ with $f_1+2f_2$ and $g_2$ with $f_2$, (\ref{eqn:anomalous-Milne-A-3d}) are the modified Milne variations
corresponding to the two different non-relativistic magnetic moment terms discussed in \cite{Jensen:2014wha}.
For $\mathcal B\neq 0$ the expressions are more bulky, but the structure is the same.
We thus find that both transformations can be realized as symmetries in a non-relativistic limit,
and we are not restricted to the linear combination with $g_1=-2g_2$ discussed in \cite{Jensen:2014wha}.
The fields were assigned trivial transformation under Milne boosts in \cite{Jensen:2014wha},
and we can understand this from our perspective as follows:
when eliminating $\Psi_-$ by its equation of motion, as done in sec.~\ref{sec:causal-spacetimes},
we found that the remaining symmetry is a combination of Milne boosts and local Galilean transformations,
precisely such that $\Psi$ is invariant.

The magnetic moment term proportional to $g_2$ was shown in sec.~2.7 of \cite{Jensen:2014aia} to reproduce the anomalous
diffeomorphisms found for $g\neq 1$ in \cite{2015PhRvD..91d5030G} as follows.
The analysis of \cite{2015PhRvD..91d5030G} should be understood as working in a fixed Milne frame.
Splitting coordinates into ($x^0$, $x^i$) as in \cite{2015PhRvD..91d5030G}, it is fixed to $v=e^\Phi\partial_0$
and we have $h=g_{ij}dx^i\otimes dx^j$.
Now the combined transformation of $v$ under a diffeomorphism generated by a vector field $X$ and a Milne boost is
$\delta v=\mathcal L_X v+\zeta$, or
\begin{align}
 \delta v^\mu&=X^\nu\partial_\nu v^\mu-v^\nu\partial_\nu X^\mu+\zeta^\mu~.
\end{align}
To compensate the change in Milne frame brought about by a diffeomorphism,
i.e.\ to keep $v^i=0$,
it has to be accompanied by a Milne boost
with $\zeta_\mu=h_{\mu\nu} v^\rho\partial_\rho X^\nu$.
The combined diffeomorphism and Milne transformations with $\zeta_\mu=h_{\mu\nu} v^\rho\partial_\rho X^\nu$
are then the symmetries discussed in \cite{2015PhRvD..91d5030G}.

As a last remark we note that, since local Galilean transformations were induced from the relativistic theory as a combination
of Milne boosts and local Lorentz transformations,
having $g\neq 1$ also changes the behavior under local Galilean transformations.

\section{Discussion}\label{sec:discussion}

We have given a procedure to construct covariant non-relativistic Lagrangians for spinor fields in general Newton-Cartan backgrounds
from relativistic parents.
As emphasized already in \cite{LevyLeblond:1967zz}, spin is tied intrinsically to the spacetime symmetries also
in the non-relativistic setting, as opposed to factoring off as internal symmetry.
For a given Newton-Cartan structure $(n,h^{-1})$, one constructs a pseudo-Riemann metric (\ref{eqn:metric-split})
and our limiting procedure gives the non-relativistic Lagrangian (\ref{eqn:NR-Lagrangian}) with the
non-relativistic Dirac/L\'evy-Leblond operator (\ref{eqn:massive-NR-Dirac}).
This Lagrangian has all desired invariances -- invariance under local Galilean transformations,
Milne boosts, U($1$) transformations and the symmetries of the Newton-Cartan structure -- and the individual objects transform covariantly.
For that result it was crucial to insist on having a non-trivial limit also for acausal Newton-Cartan spacetimes with non-vanishing $n\wedge dn$.
There is no need to identify local Galilean with Milne boosts in our construction, and under diffeomorphisms all quantities transform covariantly.
Upon gauge fixing, this setting can be specialized to recover previous results: fixing the timelike part of the inverse vielbein to $v$
links local Galilean to Milne boosts, as done in \cite{Geracie:2015dea}, and fixing a Milne frame reproduces the diffeomorphisms of \cite{2015PhRvD..91d5030G}.
The non-relativistic spinor fields have as many components as a relativistic Dirac spinor, in accordance with \cite{LevyLeblond:1967zz,Brooke:1978tr, Kunzle:1984dt}.
Half of the components are auxiliary fields and substituting them by their on-shell values retains half of the degrees of freedom as dynamical fields,
but obscures some of the symmetries. This connects our results to the language used in part of the more recent literature.

In order to have the entire set of symmetries realized without restricting the number of spacetime dimensions,
we needed to include a non-minimal coupling term in the relativistic parent with a specific strength.
This results in the non-relativistic spinor coupling with gyromagnetic ratio $g=1$.
An exception occurs for three-dimensional spacetimes.
In three dimensions, generic $g$ is possible,
but $g\neq 1$ results in modified Milne transformations, which upon gauge fixing reproduces the anomalous diffeomorphisms found in earlier approaches.
We did not specifically study the two-dimensional case, but expect that generic $g$ will be possible then without anomalous transformations.
For dimensions greater than three, our construction requires $g = 1$, unless part of the symmetries are sacrificed
along with the existence of a non-trivial limit when $n\wedge dn\neq 0$.
These results suggest a dependence on the chosen Milne frame when $g\neq 1$.

An immediate generalization of the constructions laid out in this work is to include torsion in the relativistic parent theory.
The Newton-Cartan structure $(n,h^{-1})$ can be and has been equipped with a connection as extra structure,
although somewhat implicitly.
Our construction started in the relativistic theory with the Levi-Civita connection, which gave a non-relativistic connection without
spatial torsion in the $c\rightarrow\infty$ limit. Including torsion in the parent theory would generalize this.
Another forward direction is to take our theory as starting point for an investigation of anomalies in an intrinsically non-relativistic
setting, without resorting to a relativistic parent and null reduction \cite{Jensen:2014hqa,Jain:2015jla}.
Finally, we mention the construction of supermultiplets for Newton-Cartan supergravity theories in \cite{Andringa:2013mma,Bergshoeff:2015uaa}.
Our limiting procedure may help to implement a non-relativistic limit for the fermionic fields directly in relativistic supergravities,
to obtain the transformation laws and dynamics for Newton-Cartan supergravities in a very direct way from known relativistic theories.

\section*{Acknowledgments}
It is our pleasure to thank Kristan Jensen for many helpful discussions and for comments on the manuscript.
We also thank Matthew Roberts, Kartik Prabhu and Dam Thanh Son for useful correspondence.
The work of JFF and AK was supported, in part, by the US Department of Energy under grant number DE-SC0011637.
CFU is supported by {\it Deutsche Forschungsgemeinschaft} through a research fellowship.

\bibliography{NR-reduction}
\end{document}